\documentclass[longauth]{aa} 
\usepackage[]{natbib}
\usepackage{graphicx,color}
\usepackage{txfonts}
\usepackage[colorlinks=true, citecolor=blue]{hyperref}
\usepackage[table]{xcolor}

\newcommand{\orcid}[1]{\href{https://orcid.org/#1}{\includegraphics[width=10pt]{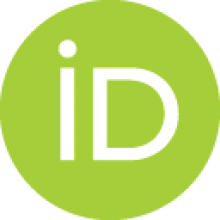}}}

\begin{document} 

\title{CAPOS: the bulge Cluster APOgee Survey \\ II.  The Intriguing "Sequoia" Globular Cluster FSR~1758}

	\author{
		Mar\'ia Romero-Colmenares\inst{1,2}\thanks{To whom correspondence should be addressed, Email: Mar\'ia Romero-Colmenares (maria.romero@uamail.cl  and/or  maria.romero21@alumnos.uda.cl)}\orcid{0000-0001-6515-5036},
		Jos\'e G. Fern\'andez-Trincado\inst{2}\thanks{in .cc to: Jos\'e G. Fern\'andez-Trincado (jose.fernandez@uda.cl and/or jfernandezt87@gmail.com)}\orcid{0000-0003-3526-5052}, 
		Doug Geisler\inst{3,4,5},
		Stefano O. Souza\inst{6},
            Sandro Villanova\inst{3},
        Pen\'elope Longa-Pe\~na\inst{1},
		Dante Minniti\inst{7,8}\orcid{0000-0002-7064-099X}, 
		Timothy C. Beers\inst{9}, 
		Cristian Moni Bidin\inst{10},
		Angeles P\'erez-Villegas\inst{11},
		Edmundo Moreno\inst{12},
		Elisa R. Garro\inst{7},
		Ian Baeza\inst{3},
		Lady Henao\inst{3}\orcid{https://orcid.org/0000-0002-2036-2944},
		Beatriz Barbuy\inst{6},
		Javier Alonso-Garc\'ia\inst{1, 13}, 
		Roger E. Cohen\inst{14}, 
		Richard R. Lane\inst{2},
		\and
		Cesar Mu\~noz\inst{4,5}
	}
	
	\authorrunning{Mar\'ia Romero-Colmenares et al.} 
	
\institute{
        Centro de Astronom\'ia (CITEVA), Universidad de Antofagasta, Av. Angamos 601, Antofagasta, Chile
	    \and 
        Instituto de Astronom\'ia y Ciencias Planetarias, Universidad de Atacama, Copayapu 485, Copiap\'o, Chile
        \and
        Departamento de Astronom\'ia, Casilla 160-C, Universidad de Concepci\'on, Concepci\'on, Chile
        \and 
         Departamento de Astronom\'ia, Universidad de La Serena, 1700000 La Serena, Chile
        \and
        Instituto de Investigaci\'on Multidisciplinario en Ciencia y Tecnolog\'ia, Universidad de La Serena. Benavente 980, La Serena, Chile
		\and
        Universidade de S\~ao Paulo, IAG, Rua do Mat\~ao 1226, Cidade Universit\'aria, S\~ao Paulo 05508-900, Brazil
        \and 
	    Depto. de Cs. F\'isicas, Facultad de Ciencias Exactas, Universidad Andr\'es Bello, Av. Fern\'andez Concha 700, Las Condes, Santiago, Chile
	    \and
	    Vatican Observatory, V00120 Vatican City State, Italy
	    \and 
	    Department of Physics and JINA Center for the Evolution of the Elements, University of Notre Dame, Notre Dame, IN 46556, USA
	    \and 
	    Instituto de Astronom\'ia, Universidad Cat\'olica del Norte, Av. Angamos 0610, Antofagasta, Chile
	    \and 
	    Instituto de Astronom\'ia, Universidad Nacional Aut\'onoma de M\'exico, Apdo. Postal 106, 22800 Ensenada, B.C., M\'exico
	    \and 
	    Instituto de Astronom\'ia, Universidad Nacional Aut\'onoma de M\'exico, Apdo. Postal 70264, M\'exico D.F., 04510, M\'exico
	    \and
	    Millennium Institute of Astrophysics , Nuncio Monse\~nor Sotero Sanz 100, Of. 104, Providencia, Santiago, Chile 
	    \and
	    Space Telescope Science Institute, 3700 San Martin Drive, Baltimore, MD 21218, USA
    }
	
	\date{Received ...; Accepted ...}
	\titlerunning{The Intriguing Globular Cluster FSR~1758}
	
	
	\abstract
	{
We present results from a study of fifteen red giant members of the intermediate-metallicity globular cluster (GC) FSR~1758 using high-resolution near-infrared spectra collected with the Apache Point Observatory Galactic Evolution Experiment II survey (APOGEE-2), obtained as part of CAPOS (the bulge Cluster APOgee Survey). Since its very recent discovery as a massive GC in the bulge region, evoking the name Sequoia, this has been an intriguing object with a highly debated origin, and initially led to the suggestion of a purported progenitor dwarf galaxy of the same name. In this work, we use new spectroscopic and astrometric data to provide additional clues to the nature of FSR~1758. Our study confirms the GC nature of FSR~1758, and as such we report for the first time the existence of the characteristic N-C anti-correlation and Al-N correlation, revealing the existence of the multiple-population phenomenon, similar to that observed in virtually all GCs. Furthermore, the presence of a population with strongly enriched aluminium makes it unlikely FSR~1758 is the remnant nucleus of a dwarf galaxy, as Al-enhanced stars are uncommon in dwarf galaxies. We find that FSR~1758 is slightly more metal rich than previously reported in the literature, with a mean metallicity [Fe/H] between $-1.43$ to $-1.36$ (depending on the adopted atmospheric parameters), and with a scatter within observational error, again pointing to its GC nature. Overall, the $\alpha$-enrichment ($\gtrsim+0.3$ dex), Fe-peak (Fe, Ni), light- (C, N), and odd-Z (Al) elements follow the trend of intermediate-metallicity GCs.   Isochrone fitting in the \textit{Gaia} bands yields an estimated age of $\sim11.6$ Gyr. We use the exquisite kinematic data, including our CAPOS radial velocities and \textit{Gaia} eDR3 proper motions, to constrain the \textit{N}-body density profile of FSR~1758, and found that it is as massive ($\sim$2.9$\pm 0.6 \times 10^{5}$ M$_{\odot}$) as NGC~6752.  We confirm a retrograde and eccentric orbit  for FSR~1758. A new examination of its dynamical properties with the \texttt{GravPot16} model favors an association with the Gaia-Enceladus-Sausage accretion event. Thus, paradoxically, the cluster that gave rise to the name of the Sequoia dwarf galaxy does not appear to belong to this specific merging event.
}
	
	\maketitle
	
\section{Introduction}
\label{section1}

\begin{table*}
	\begin{center}
		\setlength{\tabcolsep}{0.7mm}  
		\caption{Photometric, kinematic, and astrometric properties of fifteen likely members of FSR~1758.}
		\begin{tabular}{lcccccccccc}
			\hline
			\hline
			APOGEE-ID & $\alpha$ & $\delta$ & J  & H  & K$_{\rm s}$	 & $(G_{BP} - G_{RP})_{0}$ & $G_0$ & \textit{RV} & $\mu_{\alpha}\cos{}(\delta)$ & $\mu_{\delta}$ \\	
			& hh:mm:ss & dd:mm:ss  &   &   &  	 &  & & km s$^{-1}$ & mas yr$^{-1}$ & mas yr$^{-1}$ \\	
			\hline
			\hline
		{\bf  S/N $>$ 60} & & &  &  & 	 & & &  &  &  \\	
			\hline
			\hline
			2M17314970$-$3956247 & 17:31:49.71 & $-$39:56:24.8 & 10.60 &  9.57 &  9.31 &  2.51 & 13.61 &  220.9    & $-$2.994$\pm$ 0.028  &  2.511$\pm$0.019 \\
			2M17305424$-$3950496 & 17:30:54.24 & $-$39:50:49.6 & 11.13 & 10.19 &  9.93 &  2.42 & 14.04 &  227.7  & $-$3.005$\pm$ 0.027  &  2.489$\pm$0.021 \\
			2M17310731$-$3953523 & 17:31:07.32 & $-$39:53:52.4 & 11.74 & 10.82 & 10.58 &  2.35 & 14.69 &  224.5 & $-$2.863$\pm$ 0.035  &  2.526$\pm$0.023 \\
			2M17310187$-$3950066 & 17:31:01.88 & $-$39:50:06.7 & 12.07 & 11.19 & 10.91 &  2.36 & 14.97 &  228.6 & $-$2.895$\pm$ 0.039  &  2.477$\pm$0.029 \\
			2M17314263$-$3947053 & 17:31:42.64 & $-$39:47:05.4 & 12.18 & 11.35 & 11.16 &  2.24 & 14.91 &  220.7 & $-$2.983$\pm$ 0.039  &  2.529$\pm$0.030 \\
			2M17312661$-$3951342 & 17:31:26.62 & $-$39:51:34.3 & 12.20 & 11.48 & 11.24 &  2.31 & 15.29 &  226.3 & $-$2.769$\pm$ 0.040  &  2.584$\pm$0.028 \\
			2M17310738$-$3952298 & 17:31:07.39 & $-$39:52:29.9 & 12.50 & 11.65 & 11.35 &  2.24 & 15.20 &  228.2 & $-$2.850$\pm$ 0.048  &  2.405$\pm$0.033 \\
			2M17311707$-$3950382 & 17:31:17.07 & $-$39:50:38.2 & 12.73 & 11.91 & 11.70 &  2.20 & 15.54 &  226.4 & $-$2.863$\pm$ 0.049  &  2.392$\pm$0.034 \\
			2M17313165$-$3949120 & 17:31:31.65 & $-$39:49:12.1 & 12.85 & 12.11 & 11.92 &  2.03 & 15.43 &  220.2 & $-$2.863$\pm$ 0.050  &  2.690$\pm$0.035 \\
			2M17311230$-$3949493 & 17:31:12.31 & $-$39:49:49.4 & 11.90 & 11.06 & 10.75 &  2.29 & 14.91 &  223.4 & $-$2.619$\pm$ 0.038  &  2.713$\pm$0.026 \\
			2M17310618$-$3943037 & 17:31:06.18 & $-$39:43:03.8 & 13.25 & 12.48 & 12.23 &  2.14 & 15.92 &  225.0 & $-$2.777$\pm$ 0.053  &  2.526$\pm$0.041 \\
			\hline
\hline
{\bf S/N $<$ 60} & & &  &  & 	 & & &  &  &  \\	
\hline
\hline
			2M17304430$-$3952504 & 17:30:44.31 & $-$39:52:50.5 & 12.62 & 11.86 & 11.67 &  2.11 & 15.27 &  225.5 & $-$2.729$\pm$ 0.041  &  2.573$\pm$0.029 \\
			2M17310004$-$3954546 & 17:31:00.04 & $-$39:54:54.7 & 13.21 & 12.43 & 12.24 &  1.96 & 15.64 &  225.7 & $-$2.774$\pm$ 0.055  &  2.506$\pm$0.041 \\
			2M17312078$-$3948416 & 17:31:20.79 & $-$39:48:41.7 & 13.16 & 12.35 & 12.20 &  2.00 & 15.74 &  225.9 & $-$2.836$\pm$ 0.057  &  2.529$\pm$0.043 \\
			2M17305633$-$3953349 & 17:30:56.34 & $-$39:53:35.0 & 13.09 & 12.39 & 12.16 &  2.08 & 15.78 &  222.2 & $-$2.832$\pm$ 0.050  &  2.367$\pm$0.038 \\
			\hline
			\hline
		\end{tabular}  \label{Table1a}
	\end{center}
\end{table*}   

The bulge of the Milky Way (MW) is populated by a large collection ($\gtrsim$40) of  ancient globular clusters (GCs) within $\sim$3.5 kpc of the Galactic center \citep[for a detailed description, see a review by][]{Bica2016, Barbuy2018, Perez-Villegas2020}, which serves as the `fossil record' of its evolutionary history \citep[][Minniti2021a]{Recio-Blanco2017, Minniti2018a, Kundu2019, Kundu2021,  FT_Subpopulation, FT_Chemodynamics, FT_Jurassic, FT_Alluminium,FT_Dynamics, FT_NGC6723, FT_Magellanic, FT_M54, Minniti2021a}.

However, until very recently only a very small number of them have been studied with sufficient detail, including uncovering the multiple-population phenomenon  \citep[][and references therein]{Cote1999, Minniti1995, Meszaros2020, FT_NGC6522,FT_UKS1, FT_NGC6723, FT_VVCL001, Geisler2021}. Due to the high and variable extinction produced by the large amount of dust present in the Galactic plane,which highly hampers optical observations, most of these GCs have remained barely explored.

More recently, with the aid of near-infrared (NIR) photometric and high-resolution spectroscopic instruments, capable of penetrating the obscuring dust, several modern surveys have allowed us to overcome these limitations and begin to explore and scientifically exploit these fascinating objects. The \textit{Vista Variables in the Via Lactea survey} \citep[VVV:][]{Minniti2010, Saito2012, Alonso2018}, and its eXtension \citep[VVVX:][]{Minniti2018} is a key large-scale NIR photometric survey of the bulge and the surrounding disk. 

In addition to providing deep NIR photometry for all known bulge GCs, VVV/VVVX  has revealed that the GC census toward the Galactic bulge is incomplete, and more than 100 new low-luminosity GC candidates have been discovered in the bulge area \citep{Minniti2017a, Minniti2017b,  Minniti2017c, Minniti2018b, Ryu2018, Bica2018, Ramos2018, Camargo2018, Camargo2019, Minniti2019, Gran2019, Palma2019, Minniti2020, Garro2021}, and toward the Sagittarius dwarf galaxy \citep[see, e.g.,][]{Minniti2021a}. 

Complementarily, large spectroscopic surveys such as the Apache Point Observatory Galactic Evolution Experiment \citep[APOGEE;][]{Majewski2017} reveal detailed information about the chemical properties of  obscured GCs for a number of species with a variety of nucleosynthetic origins with high precision ($<$0.05 dex). Fortunately, APOGEE has helped fill in the missing link of bulge GC abundances and velocities. However, despite the large number of SDSS-IV bulge fields, the Survey only observed a handful of  bulge GCs. \citet{Masseron2019}, \citet{Meszaros2020} and \citet{Meszaros2021} present the first homogeneous study of the SDSS-IV sample of 44 GCs, of which only 2 are bona fide Main Bulge GCs according to \citet{Massari2019} and have a large enough sample of well-observed members and low enough reddening for their study. This is <4\% of the total number of bulge GCs known, a disconcertingly low value. In order to take full advantage of the wealth of astrophysical detail these key objects can provide, as complete a sample as possible is essential.

\begin{figure*}
	\begin{center}
		\includegraphics[width=160mm]{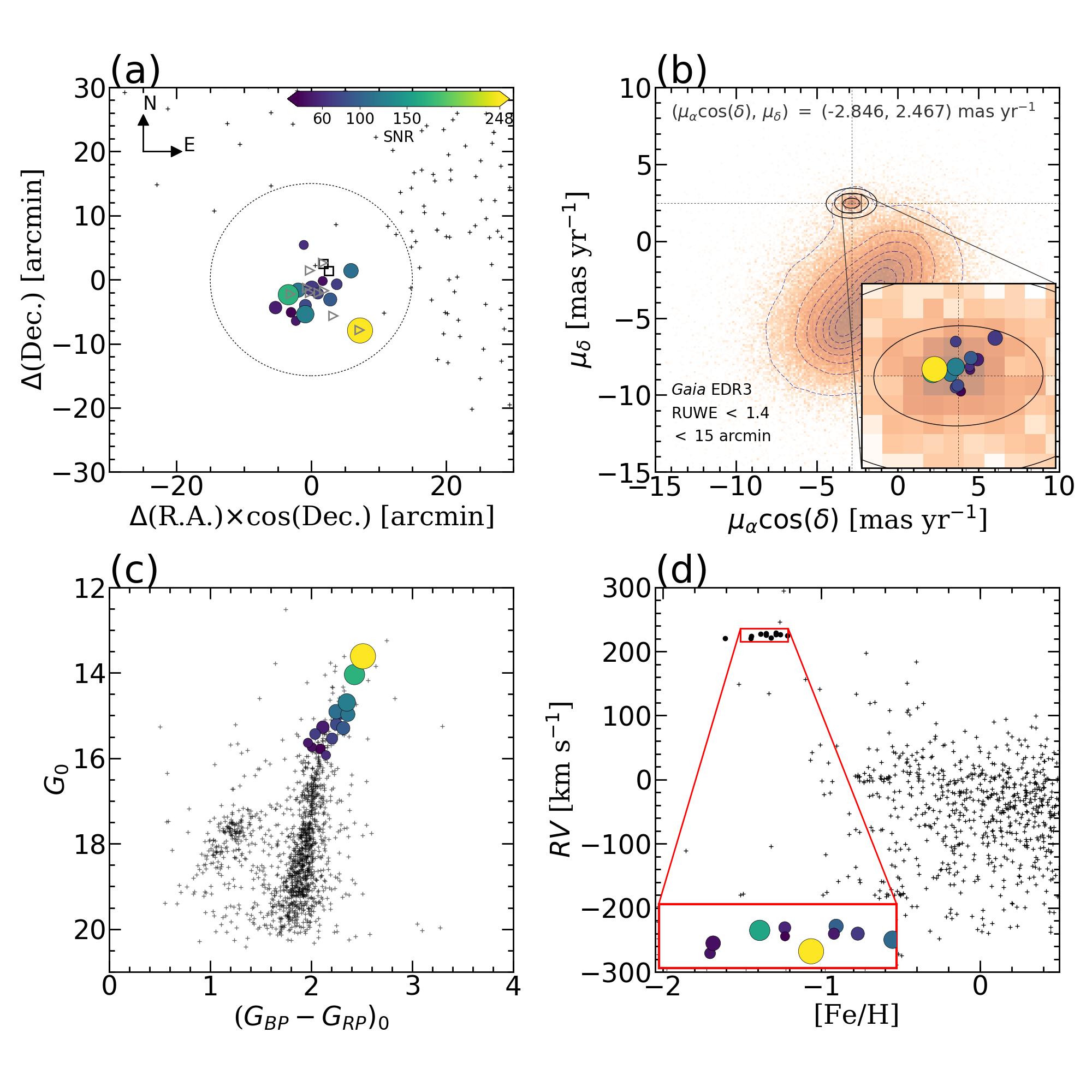}
		\caption{Properties of our FSR~1758 targets. Panel (a): Spatial position. Colour-coded symbols represent the SNR of stars with  APOGEE spectra whose sizes are proportional to their \textit{G} mag. Open black squares and grey triangles are stars analyzed in \citet{Simpson2019} and \citet{Villanova2019}, respectively. Field stars are  black crosses. A circle with 15\arcmin ~ radius is overplotted. Panel (b): The proper motion density distribution of stars located within 15\arcmin from the cluster center, with color contours refering to the Kernel Density Estimation of this sample. FSR~1758 is clearly distinguishable from the field population. The inner plot on the top-right shows a zoom-in of the cluster, with concentric ellipses showing the 1, 2, and 3$\sigma$ levels of our best fit proper motions of FSR~1758 based on \textit{Gaia} EDR3 data, whose mean values are highlighted. Symbols are the same as in panel (a). Panel (c): Colour-magnitude diagram corrected by differential reddening and extinction-corrected in the \textit{Gaia} bands of our sample and stars within 15 \arcmin. Our targets all lie along the red giant branch. Panel (d): Radial velocity  versus metallicity  of our members compared to field stars. The [Fe/H] of our targets have been determined with \texttt{BACCHUS} and photometric atmospheric parameters (see Table \ref{Table1b}), while the [Fe/H]  of field stars are from the \texttt{ASPCAP} pipeline. The red box limited by $\pm0.15$ dex and $\pm10$ km s$^{-1}$ and centered on [Fe/H]$=-1.36$ and \textit{RV} $= 225.73$ km s$^{-1}$ encloses our potential cluster members.}
		\label{Figure1}
	\end{center}
\end{figure*}

\begin{figure}
	\begin{center}
		\includegraphics[width=95 mm]{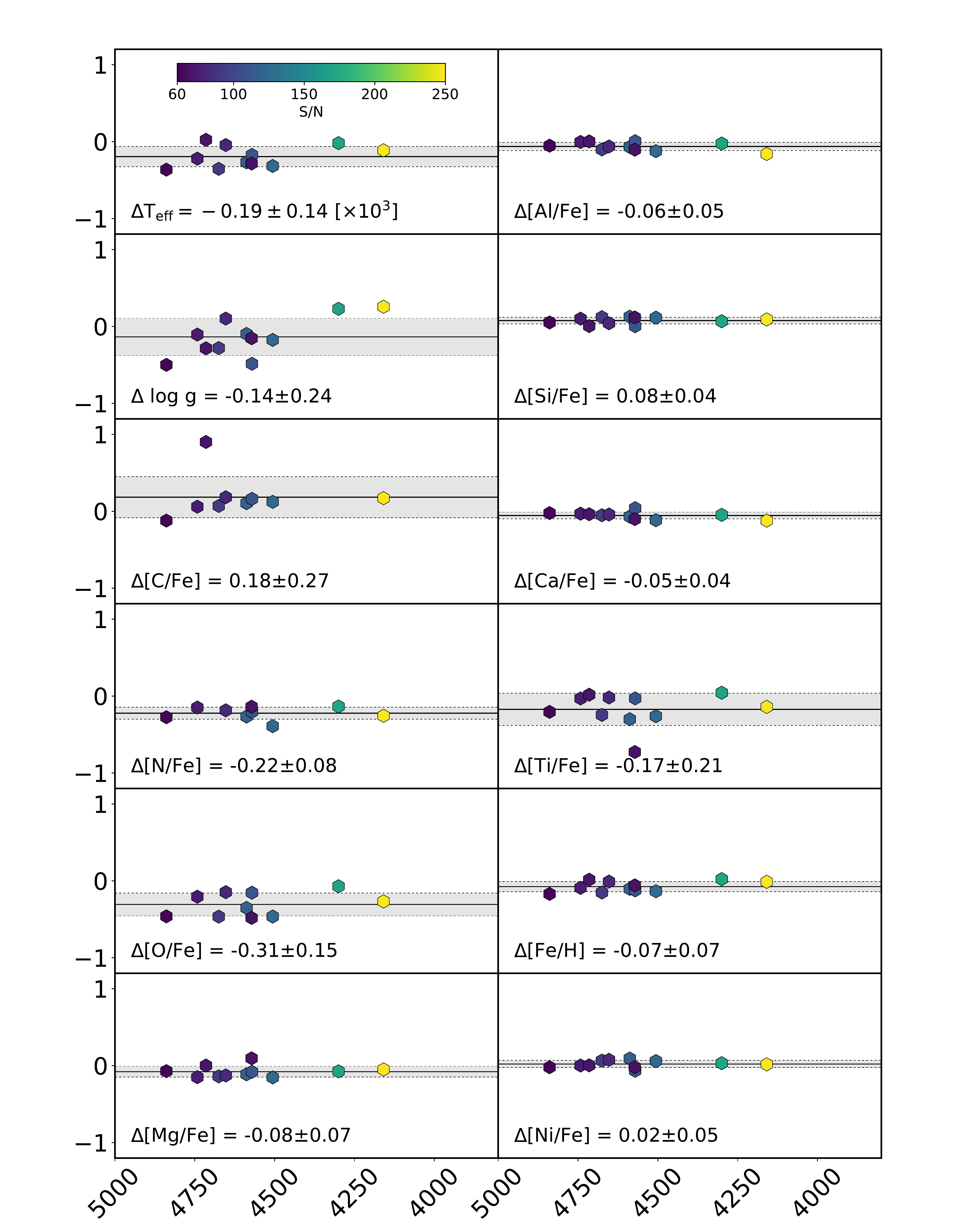}
		\caption{Differences in atmospheric parameters and elemental abundances produced by two runs adopting different effective temperatures (T$_{\rm eff}$) and surface gravities ($\log$ \textit{g}): photometric versus spectroscopic values as listed in Table \ref{Table1b}. The vertical axis refer to the $\Delta$ of every atmospheric parameter and chemical specie ([X/Fe]) analyzed in this work.  The horizontal axis refer to the {T$_{\rm eff}^{photometry}$}. The hexagonal symbols are color-coded by the S/N. The average and standard deviation around the mean of the differences is listed in each panel, and are marked by a black line and gray shadow.}
		\label{Figure8}
	\end{center}
\end{figure}

\begin{figure}
	\begin{center}
		\includegraphics[width=90mm]{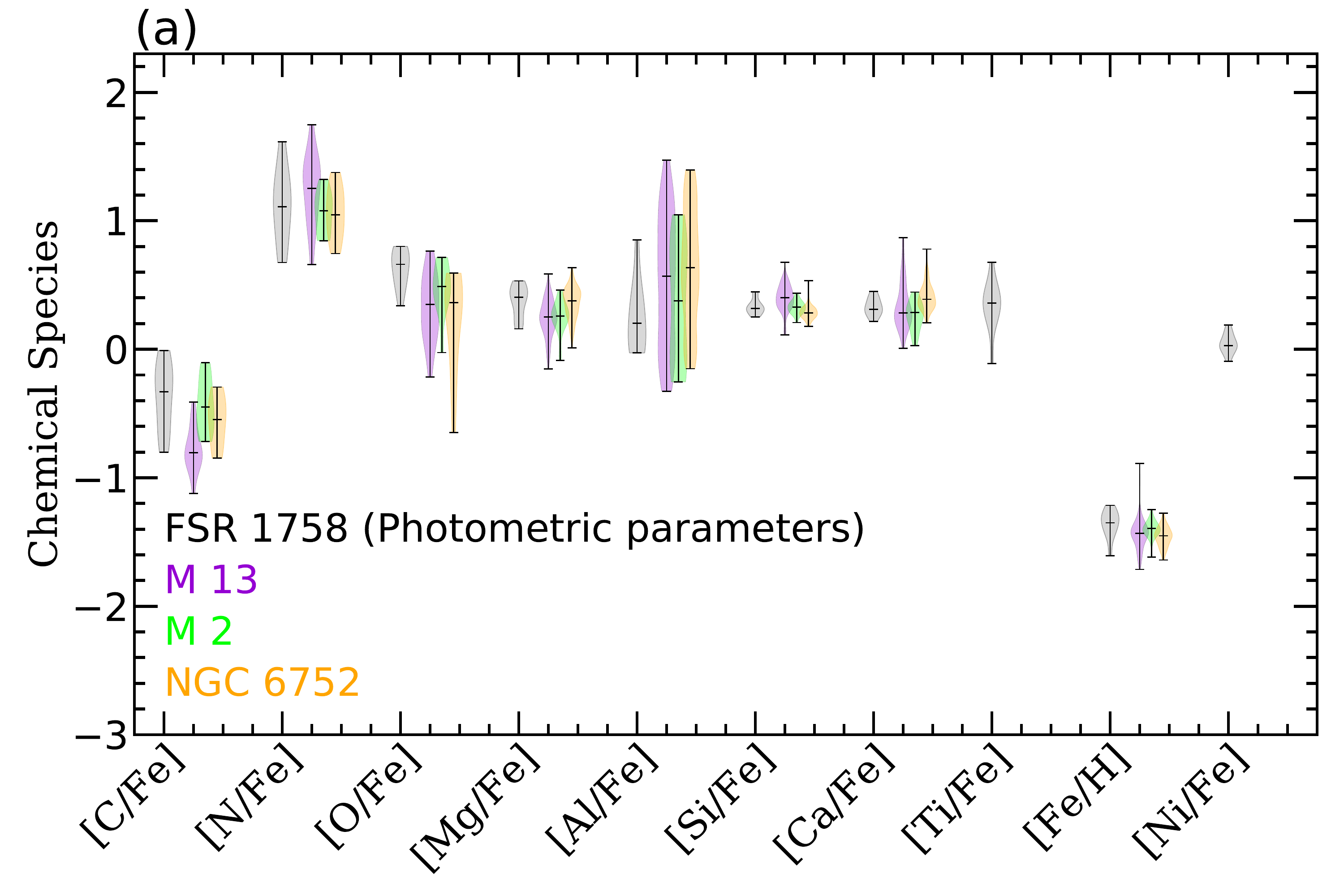}
		\includegraphics[width=90mm]{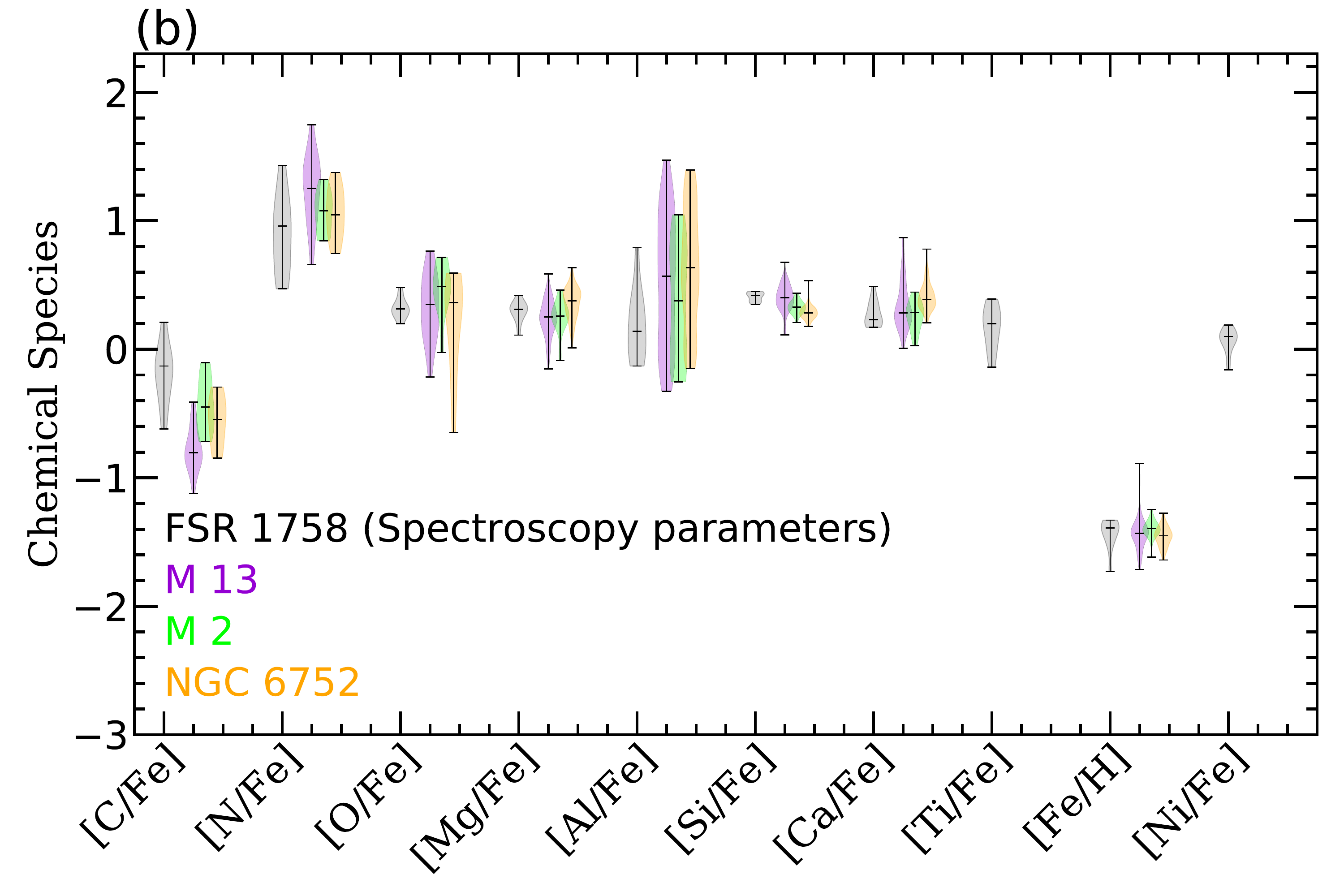}		
		\includegraphics[width=92mm]{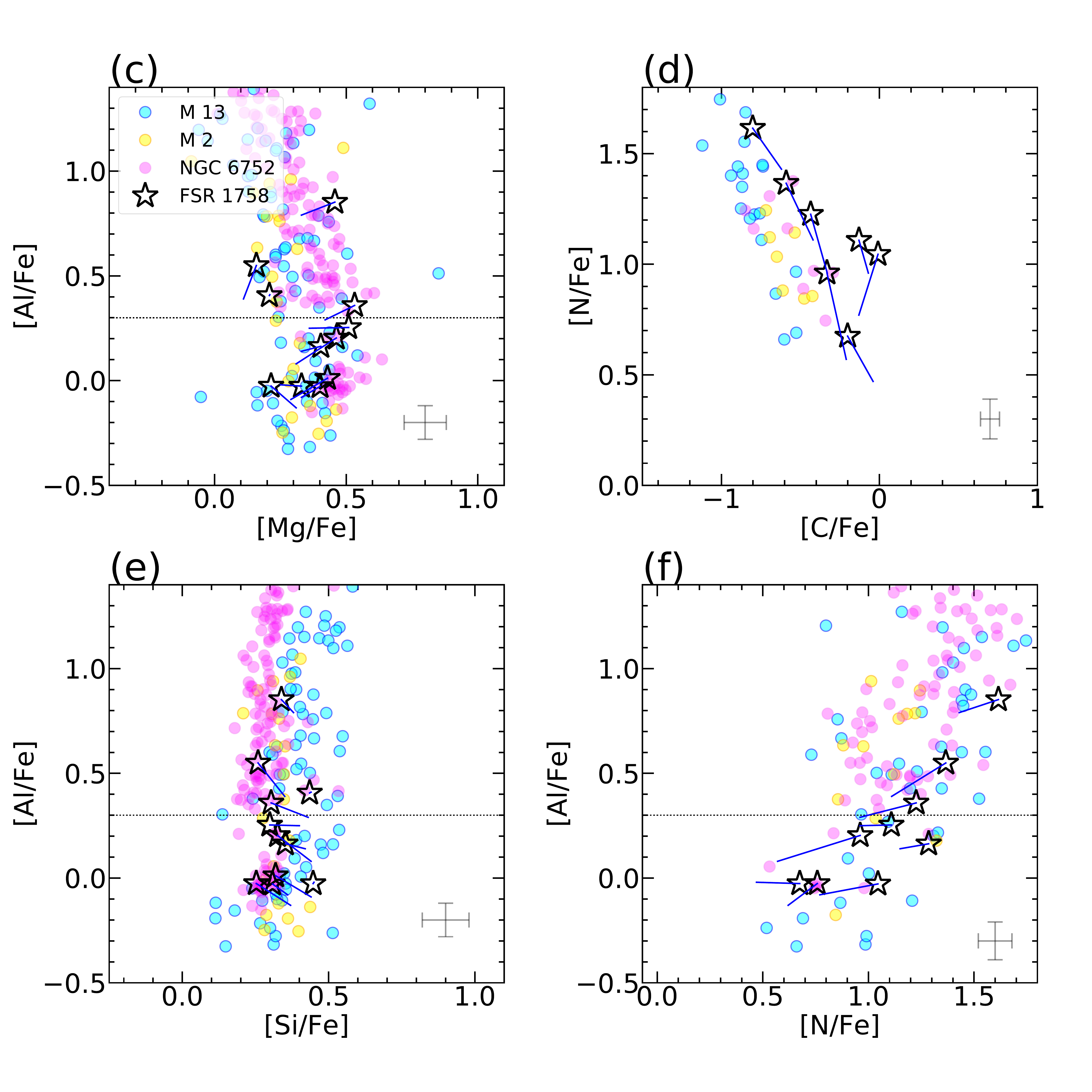}
		\caption{{\bf  Elemental abundances}. [X/Fe] and [Fe/H] abundance density estimation (violin representation) of FSR~1758 (black) compared to Galactic GCs  at similar metallicity from \citet{Meszaros2020}. Panels (a)  and (b) show our determinations by adopting photometric atmospheric parameters from photometry and spectroscopy (see Table \ref{Table1b}), respectively. Each violin representation indicates with vertical lines the median and limits of the distribution.  (c)--(f):  [Al/Fe] -- [Mg/Fe], [N/Fe] -- [C/Fe], [Al/Fe]--[Si/Fe], and [Al/Fe]--[N/Fe] distributions for GCs from \citet{Meszaros2020} at similar metallicity. The black dotted line at [Al/Fe] $=+0.3$ denotes the generalized separation of first- and second-generation stars as envisioned by \citet{Meszaros2020}. FSR~1758 stars shown `star'  symbols refers to our determinations with adopted photometry parameters, while the blue lines show the corresponding position of [X/Fe] determined from spectroscopic paramters. The typical internal error bars are also shown.}
		\label{Figure2}
	\end{center}
\end{figure}

Hence the CAPOS project, the bulge Cluster APOgee Survey (Geisler et al. 2021). The primary goal of CAPOS is to obtain detailed abundances and kinematics for as complete a sample as possible of bona fide members in bulge GCs, using the unique advantages of APOGEE in order to complement the small sample observed by SDSS-IV. 

One of the CAPOS targets is FSR~1758. It was originally catalogued as a diffuse open cluster by \citet{Froebrich2007} and \citet{Kharchenko2013}. Recently, 
\citet{Cantat-Gaudin2018} and \citet{Barba2019} re-discovered this object using a combination of data from the DECam Plane Survey \citep[DECaPS;][]{Schlafly2018} and the VVVX survey, complemented with \textit{Gaia} DR2 \citep[see][]{Brown2018}. The latter study revealed that the cluster motion is very different from nearby field stars.

Presumably located on the far side of the Galactic bulge region from the Sun at $\sim$11.5$\pm$1.0 kpc \citep{Barba2019}, FSR~1758 is as large (with a tidal radius $\sim$150 pc) as $\omega$ Cen and NGC~2419, with possibly some signatures of extra-tidal features. The GC lies in a region strongly affected by the large foreground reddening, with E(B$-$V)$> 0.7$ mags (this work), even larger than the value previously estimated, E(B$-$V)$\sim$0.37, in \citet{Barba2019}. In their re-discovery paper, \citet{Barba2019} posited that FSR~1758 could be the central part of a defunct dwarf galaxy, based on a possible halo of common proper motion stars in the surrounding regions of the cluster, or instead an unusually large GC. 

However, a subsequent spectroscopic study in the optical  \citep{Villanova2019} revealed that FSR~1758 displays no significant intrinsic metallicity spread ($\sim$0.08 dex, within the observational errors) and also only a small velocity dispersion ($<5$ km s$^{-1}$). In addition, it also possesses a  Na-O anti-correlation similar to those seen in other GCs at similar metallicity. Moreover, it is now believed that the putative extratidal stars found by \citet{Barba2019} were simply outliers in the surrounding field star proper motion distribution and likely not associated with the GC. All of this evidence supports the genuine GC nature of FSR~1758, and not the hypothesis that this is the remnant nucleus of a dwarf galaxy.  However, the dynamical properties of FSR~1758 reveal that it is not a cluster that originated in the inner Galaxy, and on the contrary  is  a halo intruder which circulates throughout the bulge in a highly eccentric, radial, and retrograde orbital configuration \citep[see, e.g.,][]{Simpson2019, Villanova2019, Yeh2020}. 

Regarding its origin, several different scenarios have been proposed for FSR~1758. \citet{Simpson2019}, based on its retrograde orbit, claimed that it could be an accreted halo GC; \citet{Villanova2019} argued that it is a genuine MW GC based on the above evidence; and \citet{Yeh2020} also supports the idea that it is likely a halo GC formed inside the MW, as the highly eccentric and retrograde orbital configuration are not uncommon among \textit{in situ} halo GCs \citep[][]{Massari2019}. On the other hand, \citet{Myeong2019} suggested that it belongs to the remnant debris of an early accretion event in the Milky Way, likely associated with a massive dwarf galaxy, similar to accretion events suggested for other GCs \citep[see, e.g.,][]{Myeong2019, Kruijssen2019}. \citet{Myeong2019} coined the name Sequoia for the putative progenitor dwarf, based on the description of FSR~1758 as a "Sequoia in the garden" given in the title of the paper by \citet{Barba2019}. Subsequent work \citep{Massari2019, Koppelman2019, FT_Subpopulation, FT_Jurassic} have shown that indeed such an event occurred, with clear kinematic and chemical evolution signatures, and the name Sequoia has stuck for the progenitor. What is not clear, and one of the issues we will address here, is whether or not this eponymous GC is indeed a member of the Sequoia dwarf. We in fact show that it was likely not a member.

In this work, we take advantage of data from the \texttt{CAPOS} survey \citep[][]{Geisler2021} to examine, for the first time in the \textit{H}-band, the chemical and kinematic properties of  FSR~1758, clarifying the characteristics of this interesting object in the dense inner regions of the MW, in particular its origin. This paper represents the second in the series on CAPOS data. In Section \ref{section2}, we describe the data. In Section \ref{fsr1758}, the potential cluster members are presented. In Section \ref{section3}, we describe the adoption of the atmospheric parameters. In Section \ref{sectionElementalAbundances}, the elemental-abundance determinations are highlighted. In Section \ref{section4}, we present our estimation of the age of the cluster by adopting a Bayesian approach, while in Section \ref{section5}, we provide a mass estimation of FSR~1758 based in the available kinematic data. In Section \ref{section6}, the origin of FSR~1758 is examined from a dynamical point of view. Summary and concluding remarks are presented in Section \ref{section7}.

\section{Data and sample}
\label{section2}

The Apache Point Observatory Galactic Evolution Experiment II survey \citep[APOGEE-2;][]{Majewski2017} is one of the internal programs of the Sloan Digital Sky Survey-IV \citep{Blanton2017, Ahumada2020} developed to provide precise radial velocities (RV $<$1 km s$^{-1}$) and detailed chemical abundances for an unprecedented large sample of giant stars, aiming to unveil the dynamical structure and chemical history of the entire MW galaxy. 

The APOGEE-2 instruments (capable of observing up to 300 objects simultaneously) are high-resolution ($R\sim22,500$), near-infrared (NIR) spectrographs \citep{Wilson2019} observing all the components of the MW (halo, disc, and bulge) from the Northern Hemisphere on the 2.5m telescope at Apache Point Observatory \citep[APO, APOGEE-2N;][]{Gunn2006} and the Southern Hemisphere on the Ir\'en\'ee du Pont 2.5m telescope \citep[][]{Bowen1973} at Las Campanas Observatory (LCO, APOGEE-2S). Each instrument records most of the \textit{H}-band (1.51$\mu$m -- 1.69$\mu$m) on three detectors, with coverage gaps between $\sim$1.58--1.59$\mu$m and $\sim$1.64--1.65$\mu$m, and with each fiber subtending a $\sim$2'' diameter on-sky field of view in the northern instrument and 1.3'' in the southern.

DR~17 is the final release of APOGEE-2 data from SDSS-III/SDSS-IV. We used our access to the internal version of DR~17, as the public version is not yet available. It includes all APOGEE data, including data taken at APO through November 2020 and at LCO through January 2021. The dual APOGEE-2 instruments have observed more than $ 650,000$ stars throughout the MW, targeting these objects with selections detailed in \citet{Zasowski2017}, \citet{Santana2021} and \citet{Beaton2021}. These papers also give a detailed overview of the targeting strategy of the APOGEE-2 survey. Spectra are reduced as described in \citet{Nidever2015}, and analyzed using the APOGEE Stellar Parameters and Chemical Abundance Pipeline \citep[ASPCAP;][]{Garcia2016}, and the libraries of synthetic spectra described in \citet{Zamora2015}. The accuracy and precision of the atmospheric parameters and chemical abundances are extensively analyzed in \citet{Holtzman2018}, \citet{Henrik2018}, and \citet{Henrik2020}, while details regarding the customised \textit{H}-band line list are fully described in \citet{Shetrone2015}, \citet{Hasselquist2016}, \citet{Cunha2017}, and \citet{Smith2021}. 

\section{FSR~1758}
\label{fsr1758}

The GC FSR~1758 was observed as part of the Contributed APOGEE-2S CNTAC\footnote{Chilean National Telescope Allocation Committee} CN2019A$-$98 program (P.I: Doug Geisler) during July 9--10, 2019 as part of the CAPOS survey \citep{Geisler2021}. 

\subsection{Proper motion selection}

The APOGEE-2S plug-plate containing the FSR~1758 cluster was centered on ($l$,$b$) $\sim$ (350, $-3.0$), and 21 of 264 science fibers were located inside a radius of $\leq 15$\arcmin from the cluster center, as shown in Figure \ref{Figure1}(a). Targets were selected on the basis of \textit{Gaia} DR2 \citep{Brown2018} proper motions within a radius of $\sim$0.7 mas yr$^{-1}$ around the nominal PMs of FSR~1758 computed by \citet{Villanova2019} from \textit{Gaia} DR2:$\mu_{\alpha}\cos(\delta)= -2.79 \pm 0.01$ mas yr$^{-1}$ and $\mu_{\delta}=2.60\pm0.01$ mas yr$^{-1}$. However, here we re-examine these values taking advantage of the improved data from \textit{Gaia} EDR3 \citep{Brown2020}. Thus, Figure \ref{Figure1}(b) shows an updated version of the PMs using \textit{Gaia} EDR3, which shows that only 15 out of the 21 originally observed ``cluster" stars share PMs similar to the nominal PMs of FSR~1758, making them likely cluster members.

As in \citet{Villanova2019}, we furthermore took potential cluster members within 15 \arcmin ~ with \textit{Gaia} EDR3 information, and having a Renormalised Unit Weight Error, \texttt{RUWE} $\leq1.4$ \citep{Lindegren2018}, in order to ensure that the selected stars were astrometrically well-behaved. Cluster members were also selected within an approximate radius of 1.0 mas yr$^{-1}$ as an initial guess for proper motion members in order to  remove field contamination from the colour-magnitude diagram (CMD). We fit one to ten Gaussians to the distribution of $\mu_{\alpha}\cos(\delta)$ and $\mu_{\delta}$ with a Gaussian Mixture Model (GMM), and measured the Akaike Information Criterion (AIC) \citep{Akaike1974} and the Bayesian Information Criterion (BIC) \citep{Schwarz1978}. Using these criteria, we then apply a GMM with only one Gaussian as this number minimized both the AIC and BIC, indicating it was the best fit. The two PM components of FSR~1758 in \texttt{Gaia} EDR3 are found to have $\mu_{\alpha}\cos(\delta)=-2.85\pm0.05$ mas yr$^{-1}$ and $\mu_{\delta}=2.47\pm0.05$ mas yr$^{-1}$--as indicated by black dashed lines in Figure \ref{Figure1}(b), which are similar to the previously estimated value from \citet[][]{Villanova2019}, and in excellent agreement with the nominal values reported in \citet{Vasiliev2021}, $\mu_{\alpha}\cos(\delta)=-2.880\pm0.026$ mas yr$^{-1}$ and $\mu_{\delta}=2.519\pm0.025$ mas yr$^{-1}$. 

Aside from PMs, stars in our sample are positioned close to the tip of the red giant branch (RGB), as shown in the \textit{Gaia} EDR3 CMD in Figure \ref{Figure1}(c). All selected stars had 2MASS \textit{H}-band  brighter than 12.5. This was required in order to achieve a nominal minimal signal-to-noise, S/N $\gtrsim$60 pixel$^{-1}$, in one plug-plate visit ($\sim$ 1 hour). Although more visits were originally planned, in the end, given weather and time allocation and airmass constraints, only 1 visit was indeed obtained. Eleven out of  the fifteen observed stars reached S/N$>60$ pixel$^{-1}$, while the remaining spectra have lower S/N, ranging from 33  to 51. In the following, we use all stars to provide reliable and precise ($< 1$ km s$^{-1}$) radial velocities for cluster member confirmation, but limit ourselves to the 11 higher S/N stars for the abundance analysis, which are highlighted in Figure \ref{Figure1}(d).

\subsection{Photometric selection}

The CMD presented in Figure \ref{Figure1}(c)  was differential reddening-corrected by using giant stars, and adopting the reddening law of \citet{Cardelli1989} and \citet{Donnell1994} and a total-to-selective absorption ratio $R_{V}=3.1$. For this purpose, we selected all RGB stars within a radius of 15 \arcmin ~ from the cluster center and that have proper motions compatible with that of FSR~1758 within 1 mas yr$^{-1}$. First, we draw a ridge line along the RGB, and for each of the selected RGB stars we calculated its distance from this line along the reddening vector. The vertical projection of this distance gives the differencial ${A_{G}}$ absorption at the position of the star, while the horizontal projection gives the differential ${E(G_{BP}-G_{RP})}$ reddening at the position of the star. After this first step, for each star of the field we selected the three nearest RGB stars, calculated the mean differential ${\rm A_{G}}$ absorption and the mean differential ${E(G_{BP}-G_{RP})}$ reddening, and finally subtracted these mean values from its ${G_{BP}-G_{RP}}$ colour and ${G}$ magnitude. We underline the fact that the number of reference stars used for the reddening correction  is a compromise between having a correction affected as little as possible by photometric random error and the highest possible spatial resolution. 

\subsection{Spectroscopic selection}

Figure \ref{Figure1}(d) reveals the \texttt{BACCHUS} [Fe/H] abundance ratios versus the radial velocity of our eleven potential cluster members compared to field stars with \texttt{ASPCAP}/APOGEE-2 [Fe/H] determinations, which have been shifted by $+$ 0.11 dex in order to minimize the systematic differences between \texttt{ASPCAP}/APOGEE-2 and \texttt{BACCHUS}, as highlighted in Appendix D in \citet[][]{Fernnadez-Trincado2020_Aluminum}. First, we note that all of our targets have very similar velocities which are extreme compared to the field star distribution. This again points to the very strong likelihood of cluster membership for all of our targets. We find a mean \textit{RV} from 15 APOGEE-2 stars of $+$224.9$\pm$ 0.7 km s$^{-1}$, which is in good agreement with \citet{Simpson2019}, $RV=+227\pm1$ km s$^{-1}$, and \citet{Villanova2019}, $RV=+226.8\pm1.6$. Other apparent sources inside this box are fore-/background stars with properties do not compatible with the cluster. 

The red box highlighted in Figure \ref{Figure1}(d) encloses the cluster members within $\pm0.15$ dex and $\pm10$ km s$^{-1}$ from the nominal mean [Fe/H]$=-1.36$ dex and \textit{RV}$= +225.73$ km s$^{-1}$ of FSR~1758, as determined in this work (see Section \ref{section5}). Note that one of our stars, 2M17314263$-$3947053, exhibits an [Fe/H] of  $-1.61$, which is $\gtrsim$ 3$\sigma$ more metal poor than the mean metallicity of the other members. For this particular case, the low-metallicity effect could be attributed to variability signature in this star \citep[see, e.g.,][for instance]{Munoz2018}, but its confirmation is beyond the scope this work. However, given the similarity of this star in \textit{RV} and PM to the kinematics/astrometry of the other stars, we consider that this is indeed a member with an outlying metallicity value. 

Table \ref{Table1a} lists the 2MASS photometry, the differential reddening corrected \textit{Gaia} photometry, and the main astrometric and kinematic properties of the observed FSR~1758 stars. 

\section{Atmospheric parameters}
\label{section3}

\subsection{The \texttt{BACCHUS} code}

Since the method of deriving stellar parameters and elemental abundances is identical to that as described in \citet{FT_Chemodynamics} and  \citet{FT_Subpopulation}, here we provide only a short overview of it. We adopted the uncalibrated T$_{\rm eff}$ and $\log$ \textit{g}, and as a first guess the uncalibrated overall metallicity, as computed by the \texttt{ASPCAP} pipeline. Secondly, we made a careful inspection of each spectrum with the \texttt{BACCHUS} code to re-derive the metallicity, broadening parameters, and chemical abundances, based on a simple line-by-line approach. The abundance determination uses the \texttt{BACCHUS} code, which relies on the radiative transfer code \texttt{Turbospectrum} \citep{Alvarez1998, Plez2012} and the \texttt{MARCS} model atmosphere grid \citep{Gustafsson2008}. The abundance of each chemical species is computed as follows: (\textit{a}) A synthesis is performed, using the full set of atomic and molecular line lists described in \citet{Shetrone2015}, \citet{Hasselquist2016}, \citet{Cunha2017}, and \citet{Smith2021}. This set of lists is internally labeled as \texttt{linelist.20170418} based on the date of creation in the format YYYYMMDD. This is used to find the local continuum level via a linear fit; (\textit{b}) Cosmic rays and telluric line rejections are performed; (\textit{c}) The local S/N is estimated; (\textit{d}) A series of flux points contributing to a given absorption line is automatically selected; and (\textit{e}) Abundances are then derived by comparing the observed spectrum with a set of convolved synthetic spectra characterized by different abundances. Then, four different abundance determination methods are used: (\textit{1}) line-profile fitting; (\textit{2}) core line-intensity comparison; (\textit{3}) global goodness-of-fit estimate; and (\textit{4}) equivalent-width comparison. Each diagnostic yields validation flags. Based on these flags, a decision tree then rejects or accepts each estimate, keeping the best-fit abundance. Here, we adopted the $\chi^2$ diagnostic as the abundance because it is the most robust. However, we store the information from the other diagnostics, including the standard deviation between all four methods.

As described in \citet{FT_Chemodynamics}, a mix of heavily CN-cycled and $\alpha$-poor \texttt{MARCS} models were used, as well as the same molecular lines adopted by \citep{Smith2013}, and employed to determine the C, N, and O abundances.  In addition, we have adopted the C, N, and O abundances that satisfy the fitting of all molecular lines consistently. We first derive $^{16}$O abundances from $^{16}$OH lines, then derive $^{12}$C from $^{12}$C$^{16}$O lines, and $^{14}$N from $^{12}$C$^{14}$N lines; the CNO abundances are derived several times initeratively to minimize the OH, CO, and CN dependences. The resulting elemental abundances are provided in the next section. 

\subsection{Photometric and Spectroscopic Parameters}

We also applied a simple approach of fixing T$_{\rm eff}$ and $\log$ \textit{g} to values determined independently of spectroscopy. In order to get T$_{\rm eff}$ and $\log$ \textit{g} from photmetry, we compared the differential reddening corrected  $G_{0}$ vs. ($G_{BP}-G_{RP}$)$_{0}$ CMD of Fig. \ref{Figure1} with isochrones obtained from the \texttt{PARSEC} database \citep{Bressan2012}. We used a preferred (see Section \ref{section4}) age of 11.6 Gyrs and a global metallicity calculated taking both [Fe/H] and $\alpha$-enhancement according to the equation of \citet{Salaris1993}. [Fe/H] and $\alpha$-enhancement were obtained from the \texttt{BACCHUS} measurements. We got the best fit to the RGB using a distance of $\sim$8.6 kpc and an absorption A$_V$=2.5. After that, we projected horizontally the position of each star until it intersected the isochrone and assumed T$_{\rm eff}$ and $\log$ \textit{g} to be the temperature and gravity of the point of the isochrones that have the same G magnitude of the star. We underline the fact that for highly reddened objects like FSR~1758 the absorption correction depends on the spectral energy distribution of  the star, i.e. on its temperature. For this reason we applied a temperature-dependent absortion correction to the isochrone. Without this trick, it is not possbile to obtain a proper fit of the RGB, expecially of the upper and cooler part. The adopted stellar parameters are listed in Table \ref{Table1b}. 

In Figure \ref{Figure8}, we compare the sensitivity to the derived atmospheric parameters, depending on the species and line in question. When the spectroscopic and photometry-based atmospheric parameters were adopted, we found discrepancies in the effective temperature of the order of $\sim18$ -- $360$ K, and surface gravity differences between $\gtrsim 0.09$ -- $0.5$ dex. The larger uncertainties are found mostly for stars in our sample with a lower S/N spectrum, and particularly for hotter ($\gtrsim4500$ K) stars. This issue does not strongly affect the derived [X/Fe] abundance ratios, but some chemical species such as carbon, nitrogen, oxygen, and titanium are more sensitive to these discrepancies in the lower S/N regime. 
It is important to note that these discrepancies do not have a large impact on our conclusions.

\begin{table*}
	\begin{small}
		\begin{center}
			\setlength{\tabcolsep}{1.mm}  
			\caption{\texttt{BACCHUS} elemental abundances of  the observed stars. The median and mean abundances, and 1$\sigma$ error are shown for the whole sample. \texttt{BACCHUS} results by adopting photometric and spectroscopic atmospheric parameters are presented in the 30 first lines of the table, which are compared to \texttt{ASPCAP} DR17 and literature results at the end of the table. All the listed elemental abundances have been scaled to the Solar reference from \citet{Asplund2005}.}
			\begin{tabular}{lcccccrrrrrrrrrr}
				\hline
				\hline
				APOGEE-ID & S/N &     [M/H] & T$_{\rm eff}$ & $\log$ \textit{g}  &   $\xi_{t}$   & [C/Fe] & [N/Fe] & [O/Fe] & [Mg/Fe]  & [Al/Fe] & [Si/Fe] & [Ca/Fe] & [Ti/Fe]  & [Fe/H]  & [Ni/Fe] \\
				& pixel$^{-1}$ &     & K & cgs &   km s$^{-1}$   &  &  &  &   & &  & & &   &  \\
				\hline
				\hline
			    { Photometry	}		&  &     &  &  &    &  &  &  &   & &  & & &   &  \\
\hline		
\hline			
2M17314970$-$3956247 & 250 & $-1.31$ & 4159  & 0.53 &  2.69  &  $-$0.59  &  $+$1.36 &   $+$0.51 & $+$0.15 & $+$0.54 &  $+$0.25 &  $+$0.34 &    $+$0.34 & $-$1.32 &    $+$0.02 \\
2M17305424$-$3950496 &  172 & $-1.34$ & 4300  & 0.80 &  2.07  &  ...              &  $+$1.28 &  $+$0.33 & $+$0.40 & $+$0.16 &  $+$0.35 &  $+$0.22 &    $+$0.22 & $-$1.38 &    $+$0.03 \\
2M17310731$-$3953523 &  126 & $-1.39$ & 4506  & 1.18 &  1.71  &  $-$0.33  &  $+$0.96 &  $+$0.77 & $+$0.46 & $+$0.20 &  $+$0.32 &  $+$0.31 &    $+$0.45 & $-$1.22 &    $+$0.01 \\
2M17310187$-$3950066 &  118 & $-1.38$ & 4588  & 1.33 &  1.90  &  $-$0.43  &  $+$1.22 &  $+$0.64 & $+$0.53 & $+$0.35 &  $+$0.30 &  $+$0.26 &    $+$0.36 & $-$1.29 &    $+$0.01 \\
2M17314263$-$3947053 &  110 & $-1.70$ & 4571  & 1.29 &  1.35  &  $-$0.20  &  $+$0.67 &  $+$0.56 & $+$0.33 & $-$0.02 &   $+$0.44 &  $+$0.45 &    $-$0.11 & $-$1.61 & $-$0.09 \\
2M17312661$-$3951342 &    93 &  $-1.41$ & 4675  & 1.49 &  1.79  &  $-$0.19  &  ...              &  $+$0.78 & $+$0.42 & $+$0.01 &  $+$0.31 &  $+$0.41 &    $+$0.46 & $-$1.26 &    $+$0.06 \\
2M17310738$-$3952298 &    82 &  $-1.31$ & 4653  & 1.45 &  1.68  &  $-$0.80  &  $+$1.61 &  $+$0.46 & $+$0.45 & $+$0.85 &  $+$0.33 &  $+$0.31 &    $+$0.39 & $-$1.35 &    $+$0.04 \\
2M17311707$-$3950382 &    75 & $-1.34$ & 4742  & 1.62 &  1.81  &  $-$0.12  &  $+$1.10 &   $+$0.68 & $+$0.50 & $+$0.25 &  $+$0.29 &  $+$0.36 &    $+$0.42 & $-$1.29 &    $+$0.19 \\
2M17313165$-$3949120 &    71 & $-1.43$ & 4715  & 1.57 &  1.47  &  $-$0.68   &  ...  &  ...  &  $+$0.20 &  $+$0.41 & $+$0.43 & $+$0.27 &  $+$0.28 &    $-$1.45 &  $+$0.13 \\
2M17311230$-$3949493 &    69 & $-1.58$ & 4572  & 1.30 &  1.64  &  ...               &  $+$0.75 &  $+$0.68 & $+$0.21 & $-$0.02 &  $+$0.25 &  $+$0.30 &    $+$0.68 & $-$1.44 &    $+$0.12 \\
2M17310618$-$3943037 &    61 & $-1.54$ & 4839  & 1.81 &  1.61  &  $-$0.00  &  $+$1.04  &  $+$0.80 & $+$0.40 & $-$0.02 &  $+$0.30 &  $+$0.39 &    $+$0.25 & $-$1.35 &     $-$0.04 \\				
\hline
Median              &  ... &  ... &  ... &  ...  &  ... & $-0.33$  & $+$1.11   & $+$0.66  & $+$0.40 & $+$0.20  & $+$0.32  & $+$0.31  & $+$0.36  & $-1.35$  &  $+$0.03  \\ 
\rowcolor{pink}
\textcolor{black}{\bf Mean    }              &  ... &  ... &  ... &  ...  &  ... & \textcolor{black}{\bf $-$0.38}  &  \textcolor{black}{\bf $+$1.12} & \textcolor{black}{\bf  $+$0.63}  & \textcolor{black}{\bf  $+$0.37} & \textcolor{black}{\bf  $+$0.25}  & \textcolor{black}{\bf  $+$0.33}  &\textcolor{black}{\bf  $+$0.33}  &\textcolor{black}{\bf  $+$0.34} & \textcolor{black}{\bf  $-$1.36}  & \textcolor{black}{\bf  $+$0.04} \\ 
$1\sigma$       &  ... &  ... &  ... &  ...  & ...  &    0.26      &  0.26  & 0.15 &  0.13 &  0.24 & 0.05  & 0.07  & 0.11  & 0.08         &  0.07  \\   
\hline
\hline
					{ Spectroscopy		}	&  &     &  &  &    &  &  &  &   & &  & & &   &  \\
				\hline
				\hline
				2M17314970$-$3956247 &  250  &  $-$1.31 &  4047  &  0.79 &  2.66  & $-$0.42 &   $+$1.11 &   $+$0.25 &   $+$0.11 &     $+$0.39 &   $+$0.35 &  $+$0.22 &    $+$0.20 &  $-$1.33  &    $+$0.04 \\
				2M17305424$-$3950496 &  172  &  $-$1.34 &  4281  &  1.03 &  1.92  &     ...         &   $+$1.15 &   $+$0.27 &   $+$0.33 &     $+$0.14 &   $+$0.42 &  $+$0.17 &    $+$0.26 &  $-$1.36  &    $+$0.06 \\
				2M17310731$-$3953523 &  126  &  $-$1.39 &  4190  &  1.01 &  1.69  & $-$0.21 &   $+$0.57 &   $+$0.31 &   $+$0.31 &     $+$0.08 &   $+$0.44 &  $+$0.19 &    $+$0.19 &  $-$1.35  &    $+$0.07 \\
				2M17310187$-$3950066 &  118  &  $-$1.38 &  4320  &  1.23 &  1.72  & $-$0.33 &   $+$0.96 &   $+$0.29 &   $+$0.42 &     $+$0.29 &   $+$0.43 &  $+$0.19 &    $+$0.06 &  $-$1.39  &    $+$0.10 \\
				2M17314263$-$3947053 &  110  &  $-$1.70 &  4398  &  0.81 &  1.63  & $-$0.04 &   $+$0.47 &   $+$0.41 &   $+$0.25 &     $-$0.02 &   $+$0.45 &  $+$0.49 & $-$0.14 &  $-$1.73  & $-$0.16\\
				2M17312661$-$3951342 &   93  &  $-$1.41 &  4322  &  1.22 &  1.58  & $-$0.13   &    ...             &   $+$0.32 &   $+$0.29 &     $-$0.09 &   $+$0.44 &  $+$0.36 &    $+$0.22 &  $-$1.41  &    $+$0.13 \\
				2M17310738$-$3952298 &   82  &  $-$1.31 &  4609  &  1.56 &  1.62  & $-$0.62   &   $+$1.43 &   $+$0.32 &   $+$0.33 &     $+$0.79 &   $+$0.38 &  $+$0.27 &    $+$0.38 &  $-$1.36  &    $+$0.12 \\
				2M17311707$-$3950382 &   75  &  $-$1.34 &  4522  &  1.52 &  1.83  & $-$0.07   &   $+$0.96 &   $+$0.48 &   $+$0.36 &     $+$0.25 &   $+$0.40 &  $+$0.33 &    $+$0.39 &  $-$1.38  &    $+$0.19 \\
				2M17313165$-$3949120 &   71  &  $-$1.43 &  4740  &  1.29 &  1.47  & $+$0.21  &    ...             &   ...              &   $+$0.21 &     $+$0.41 &   $+$0.44 &  $+$0.23 &    $+$0.30 &  $-$1.43  &    $+$0.13 \\
				2M17311230$-$3949493 &   69  &  $-$1.58 &  4289  &  1.15 &  0.85  &     ...           &   $+$0.62 &   $+$0.20 &   $+$0.31 &     $-$0.13 &   $+$0.37 &  $+$0.20 & $-$0.05 &  $-$1.50  &    $+$0.10 \\
				2M17310618$-$3943037 &   61 &  $-$1.54 &  4476  &  1.32 &  1.49  & $-$0.13    &   $+$0.77 &   $+$0.34 &   $+$0.33 &     $-$0.08 &   $+$0.36 &  $+$0.37 & $+$0.05 &  $-$1.52  & $-$0.06 \\	
				\hline
				Median              & ... & ...  &  ... & ...   & ...  & $-0.13$  &  $+$0.96  & $+$0.32 & $+$0.31 &  $+$0.14  &$+$0.42 & $+$0.23  & $+$0.20  & $-1.39$ & $+$0.10 \\ 
				\rowcolor{pink}
			\textcolor{black}{\bf Mean}                  & ...  & ...  & ...  & ...   & ...  & \textcolor{black}{\bf $-$0.19}  & \textcolor{black}{ \bf $+$0.89} &  \textcolor{black}{\bf $+$0.32} & \textcolor{black}{\bf $+$0.29}  & \textcolor{black}{\bf $+$0.18} & \textcolor{black}{\bf $+$0.41}  & \textcolor{black}{\bf $+$0.27}  & \textcolor{black}{\bf $+$0.17} & \textcolor{black}{\bf  $-$1.43}  & \textcolor{black}{\bf  $+$0.07} \\ 
				$1\sigma$       &  ... & ...  & ...  & ...   &  ... &    0.17      &  0.28  & 0.06 & 0.05 & 0.24 & 0.04   & 0.09  & 0.16  & 0.08      & 0.07 \\   
				\hline
				\hline	
				{ \texttt{ASPCAP} DR17}			&  &     &  &  &    &  &  &  &   & &  & & &   &  \\
				\hline
				\hline					
Median              & ...  & ...  & ...  &  ...  &  ... & $-0.41$  & $+$0.89   & $+$0.35  & $+$0.34 & $+$0.21  & $+$0.27  & $+$0.29  & $+$0.02  & $-1.38$  & $-0.02$  \\ 
\rowcolor{pink}
\textcolor{black}{\bf  Mean}                  &  ... & ...  &  ... &  ...  &  ... & \textcolor{black}{\bf $-$0.41}  & \textcolor{black}{\bf $+$0.71}   & \textcolor{black}{\bf $+$0.31}  & \textcolor{black}{\bf $+$0.29} & \textcolor{black}{\bf $+$0.18}  & \textcolor{black}{\bf $+$0.27}  & \textcolor{black}{\bf $+$0.27}  & \textcolor{black}{\bf $-$0.01} & \textcolor{black}{\bf $-$1.43}  & \textcolor{black}{\bf $-$0.01}  \\ 
$1\sigma$       &  ... & ...  & ...  & ...   &  ... & $0.12$   & 0.34   & 0.11  & 0.07 & 0.29  & 0.04  & 0.05  & 0.15   & 0.09    & 0.05 \\   
\hline
\hline
\citet{Villanova2019}			&  &     &  &  &    &  &  &  &   & &  & & &   &  \\
\hline
\hline				
 Median                  &  ... &  ... &  ... &  ...  &  ... & ... &  ... & $+$0.37  &  $+$0.67 &  $+$0.44 & ...  &  $+$0.37  &  $+$0.33 & $-$1.49 &  $-$0.10  \\ 
\rowcolor{pink}
\textcolor{black}{\bf Mean}                  &  ... &  ... &  ... &  ...  &  ... & ... &  ... & \textcolor{black}{\bf $+$0.32}  & \textcolor{black}{\bf $+$0.67} & \textcolor{black}{\bf $+$0.43}  & ...  & \textcolor{black}{\bf $+$0.35}  & \textcolor{black}{\bf $+$0.31} & \textcolor{black}{\bf $-$1.53}  & \textcolor{black}{\bf  $-$0.11}  \\ 
 $1\sigma$                    &  ... &  ... &  ... &  ...  &  ... & ... &  ... & 0.08  &  0.09 &  0.13 & ...  &  0.08  &  0.06 & 0.06 &  0.05  \\ 
\hline
\hline
			\end{tabular}  \label{Table1b}
		\end{center}
	\end{small}
\end{table*}   

 \section{Elemental abundances}
 \label{sectionElementalAbundances}
 
	We limit our discussion to [X/Fe] and [Fe/H] abundance ratios obtained from a line-by-line inspection of the APOGEE-2 spectra with the \texttt{BACCHUS} pipeline \citep{Masseron2016} -- independent of \texttt{ASPCAP}, in order to reduce a number of problems produced by the blind \texttt{ASPCAP} processing in lower-metallicity cluster stars \citep[see, e.g.,][]{Masseron2019, Meszaros2020}.
 
Ten chemical species were investigated from the APOGEE-2 spectra, including the Fe-peak (Fe, Ni), $\alpha$- (O, Mg, Si, Ca, Ti),  light- (C, N), and odd-Z (Al) elements. Our study provides complementary information on elements not accessible from \cite{Villanova2019}, therefore C, N, and Si are reported for the first time. The \texttt{BACCHUS} elemental abundances are listed in Table \ref{Table1b} highlighting the mean value of each chemical specie in FSR~1758 from different methods. Internal errors were derived using sensitivities of the abundances to variations in the adopted atmospheric parameters, and the internal uncertainties in each parameter as estimated in \citet{FT_Jurassic}.

This study now places our chemical understanding of FSR~1758 alongside that of other well-studied intermediate-metallicity GCs. Overall, the results for FSR~1758  are in agreement with other GCs of similar metallicity for all species, as shown in Figure \ref{Figure2}(a).

\subsection{The iron-peak elements: Fe, and Ni}

We found a mean metallicity [Fe/H] $=-1.36$ and [Fe/H] $=-1.43$ depending on the adopted atmospheric parameters. This difference in [Fe/H] is smaller that the internal 1$\sigma$ deviation, 0.08 dex, as listed in Table \ref{Table1b}. Furthermore, the results listed in Table \ref{Table1b} reveal there is no significant metallicity spread, which confirms that FSR~1758 is consistent with other GCs at similar metallicity, as seen in Figure \ref{Figure2}(a). However, it is important to note that our mean metallicity is on average $\sim0.10$--$0.17$ dex higher than the mean  metallicity ([Fe/H]$= -1.53$) determined from optical spectra \citep[][]{Villanova2019}, and $\sim$0.07 higher than \texttt{ASPCAP} DR17 determination when compared to our determinations by adopting photometric parameters. These differences are highlighted in Table \ref{Table1b}.

The nature of this discrepancy could be attributed to several different factors. Unfortunately, there are no common stars between these studies. However, we list some possible causes. Different reference solar abundances or atmospheric models could be responsible. Another possible source of systematic differences could be the result of NLTE and/or 3D effects, which are currently not modeled when fitting the APOGEE spectra \citep[see, e.g.,][]{Masseron2019, Meszaros2020}.

\begin{figure}
	\begin{center}
		\includegraphics[width=95 mm]{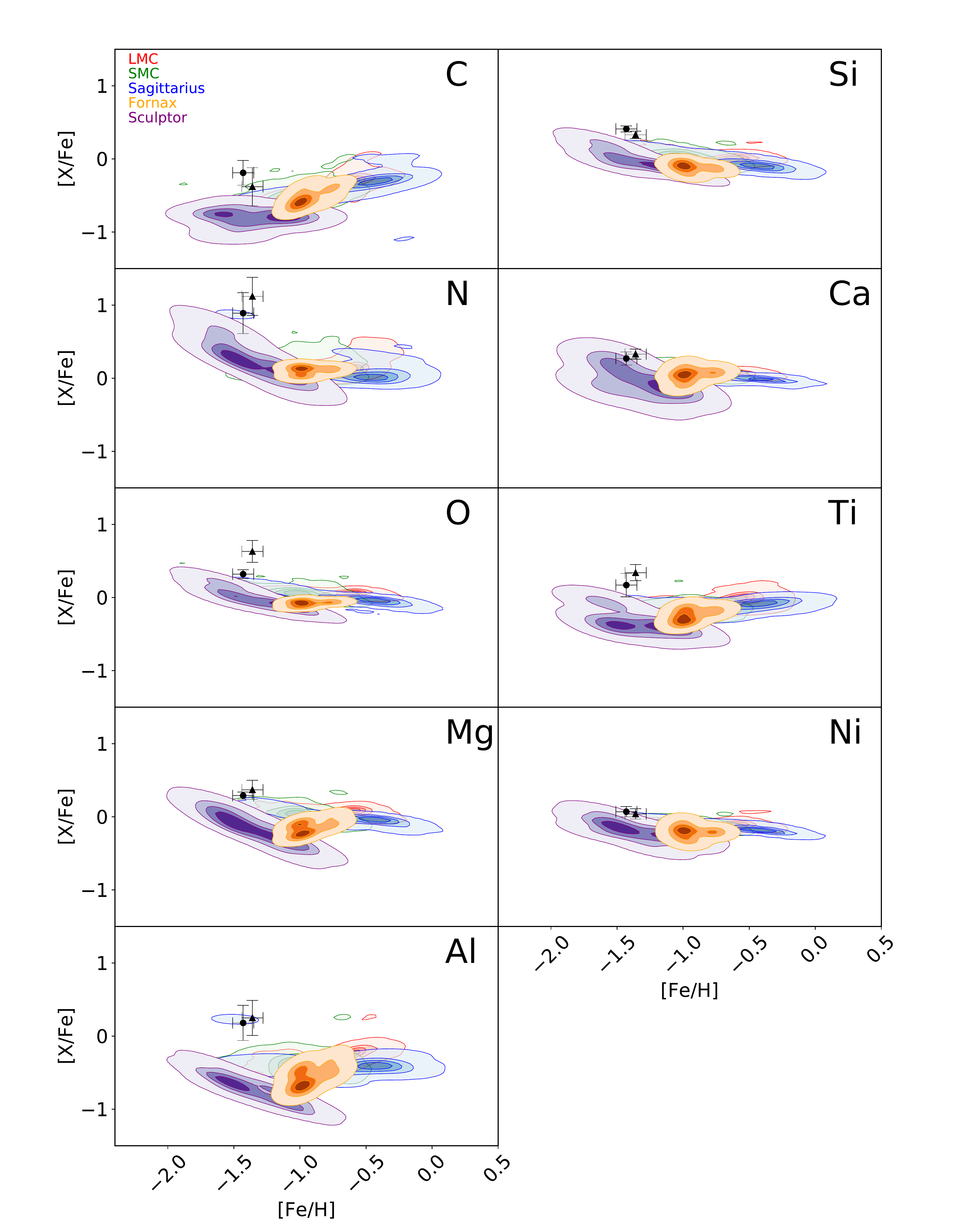}
		\caption{Kernel Density Estimation (KDE) models of [X/Fe] versus [Fe/H] for star in extragalactic environments, taken from the compilation of stars from \citet{Helmi2018} with \texttt{ASPCAP} DR17 determinations. The black symbols show our \texttt{BACCHUS} elemental abundances determined from spectroscopic (dot symbol) and photometric (triangle symbol) atmospheric parameters.}
		\label{Figure9}
	\end{center}
\end{figure}

As far as the other iron-peak element we studied, nickel (Ni) is on average slightly over-solar ([Ni/Fe]$=$ $+0.04$ -- $+0.07$) with a relatively small star-to-star [Ni/Fe] spread, $< $0.07 dex. We find a $\langle$[Ni/Fe]$\rangle$ slightly higher ($\gtrsim0.15$ dex) than [Ni/Fe] abundance ratios determined in \citet{Villanova2019}. The measured [Ni/Fe] abundance ratio in FSR ~1758 is at a similar level to that observed in extragalactic environments at similar metallicity as seen in Figure \ref{Figure9}.

\subsection{The odd-Z elements: Al}

We find median values for [Al/Fe] compatible with other Galactic GCs at similar metallicity, as shown in Figure \ref{Figure2}(b). The sample has an observed star-to-star spread in [Al/Fe] of $\gtrsim +0.8$ dex, which greatly exceeds the observational uncertainties. There is at most a weak Al-Mg anti-correlation, possibly showing the signs of a moderate Mg-Al cycle in FSR~1758, common to most intermediate to metal-poor GCs in the Galaxy \citep{Masseron2019, Meszaros2020, Meszaros2021}. Inspection of the [Si/Fe] ratio as a function of [Al/Fe] for the few stars with reliable Si measurements, as shown in Figure \ref{Figure2}(d), reveal that there is not a statistically significant correlation between Al and Si, as will be expected from the Mg-Al cycle, as Al-rich stars also present enrichment in Si \citep[see, e.g.,][]{Yong2005, Carretta2009}.  Thus, the observed Si-Al trend in Figure \ref{Figure2}(d) explains the apparent weak Al-Mg anti-correlation seen in Figure \ref{Figure2}(a).

Figure \ref{Figure2}(b:e) reveals that, by adopting the rough limit of [Al/Fe]$=+0.3$ to separate the so-called first- and second-generation stars, as suggested in \citet[][]{Meszaros2020, Meszaros2021}, there are two apparent groups of stars in FSR~1758. One of these groups, dominated by first-generation stars, exhibits low Al and (slight) Mg enrichment with intermediate N enrichment ([N/Fe]$\lesssim$ 1); while the second group, dominated by second-generation stars, shows high Al and N enrichment simultaneous with depleted C. Therefore, we conclude that FSR~1758 does indeed host multiple populations common to virtually all GCs, and that in addition to the Na-O anti-correlation identified in \citet{Villanova2019}, we also identified the signatures of N-C anti-correlation typical of GCs at similar metallicity. In particular, the presence of a population enriched in [Al/Fe] ($>+0.5$) in FSR~1758 rules out the possible scenario of FSR~1758 being the nucleus of an accreted dwarf galaxy, since large enrichment in Al has not been observed in dwarf galaxy stellar populations \citep[see, e.g.,][]{Shetrone2003, Hasselquist2017, Hasselquist2019}.

\begin{figure*}
	\begin{center}
		\includegraphics[width=90mm]{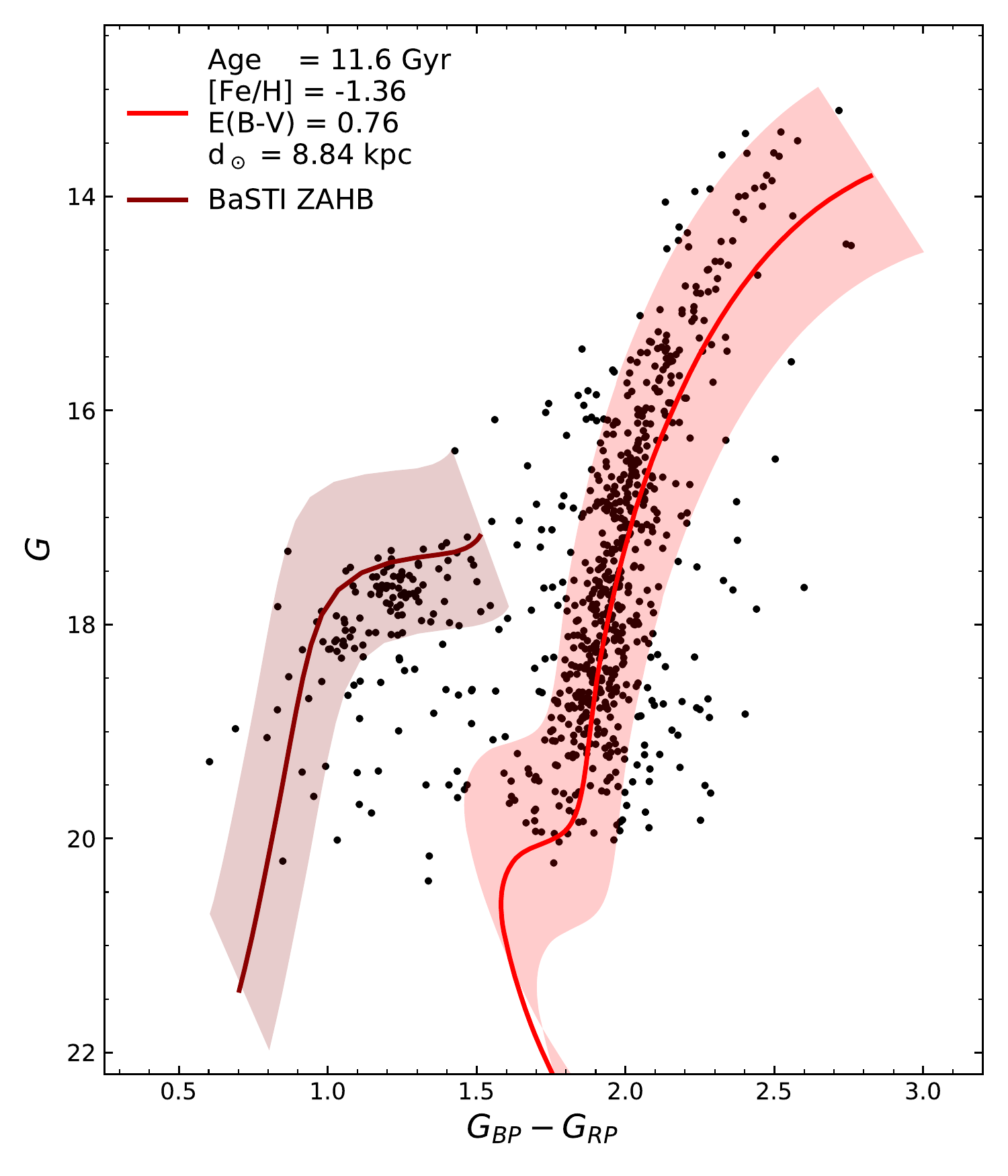}\includegraphics[width=93mm]{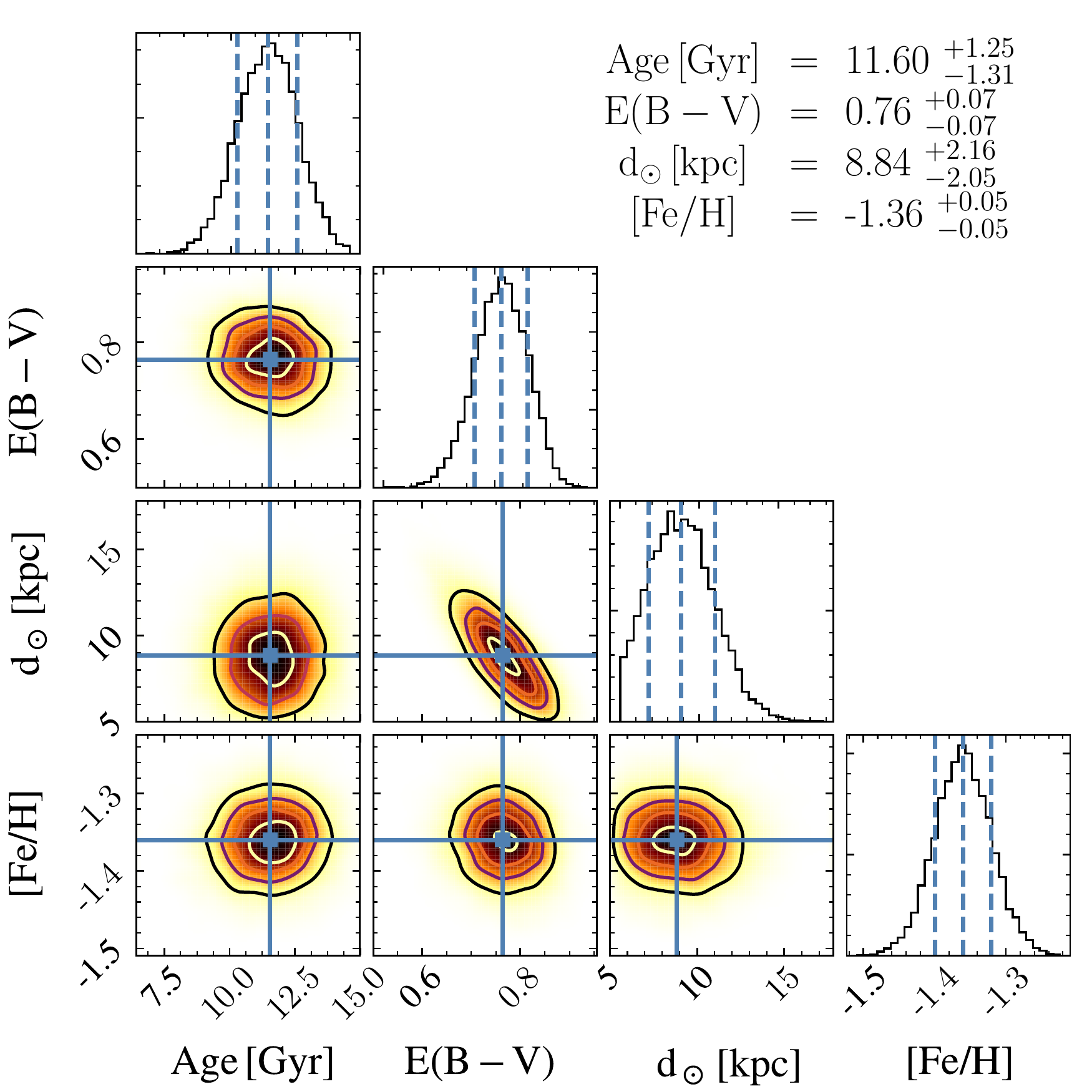}
		\caption{Age estimation of FSR~1758. Left panel: The best isochrone fit in the \textit{Gaia} bands CMD using DSED models for likely cluster members within 10 arcmin, where the red and brown lines show the most probable solutions, and the shadowed regions indicate the solutions within $1\sigma$. Right panel: The posterior distributions of the indicated quantities.}
		\label{Figure4}
	\end{center}
\end{figure*}

\subsection{The light-elements: C and N}

Figures \ref{Figure2}(c) and (e) reveal a clear N-C anti-correlation, Al-N correlation, and a large N spread ($>+0.9$ dex). We conclude that the intra-GC abundance variations reported in  Figures \ref{Figure2}(c) and (e) are indicative of the presence of multiple populations in FSR~1758. It is the first time that the presence of a spread in C and N has been established in FSR~1758.

\subsection{The $\alpha$-elements: O, Mg, Si, Ca and Ti}

Figure \ref{Figure2}(a) shows that the $\alpha$ elements in FSR~1758 follow the same trend as other GCs at similar metallicity, again confirming the genuine GC nature of FSR~1758, as originally suggested by \citet{Villanova2019}. However, as can be appreciated in Figure \ref{Figure9}, the [Ca/Fe] abundance ratio appears to be at the same level as that seen in extragalatic environments.

Finally, Figure \ref{Figure9} reveals that all the chemical species (except [Ni/Fe] and [Ca/Fe]) in our sample depart from the typical  level of extragalactic environments, making FSR~1758 unlikely associated with the nucleus of a dwarf galaxy as previously hipothetised in \citep{Barba2019}. We conclude that the measured elemental abundance ratios and anti-correlation/correlation features support the genuine GC nature of FSR~1758. 

\section{Age}
\label{section4}

We estimate the age of FSR~1758 through isochrone fitting. To accomplish this, we employ the \texttt{SIRIUS} code \citep{Souza2020}, which applies a Bayesian statistical inference based on the Markov Chain Monte Carlo method. For the isochrone fitting in the RGB region, we adopt the Darthmouth Stellar Evolutionary Database \citep[DSED:][]{Dotter2008} with an $\alpha$-enhancement of $+0.4$ and canonical helium models of Zero-Age Horizontal Branch (ZAHB) from the Bag of Stellar Tracks and Isochrones \citep[BaSTI:][]{Pietrinferni2006}.

Since we do not have the entire CMD available, in particular the turn-off region, we imposed Gaussian distribution priors for the metallicity of [Fe/H]$\sim-1.36$ (see Section \ref{sectionElementalAbundances}), based on the mean determination from this work, and by assuming an error of $\sim$0.05 dex. Metallicity plays a key role in the morphology of stellar evolutionary models, mainly in temperature (color). Lower metalicities tend to higher temperatures (blue region, see fig. 2 in \citet{Souza2020}). Since the evolutionary stage available in our case is restricted to RGB, we imposed a Gaussian prior in metallicity to limit the permitted values.

Figure \ref{Figure4} presents the best isochrone fitting of the \textit{Gaia} EDR3 CMD. Our fit provides a reasonable solution both in the overplotted isochrone (left panel) and the posterior distributions of the corner plot (right panel).  As the best determination to represent the distributions, we adopted the median as the most probable value and the uncertainties calculated from 16$^{\rm th}$ and 84$^{\rm th}$ percentiles. A close inspection of the \textit{Gaia} EDR3 CMD reveals that a lower heliocentric distance of $\sim$8.84$^{+2.16}_{-2.05}$ kpc better fits the horizontal and RGB simultaneously, with evident large uncertainties likely due to false-positive members located in the red side of the CMD, yielding an age estimate of $\sim11.6^{+1.25}_{-1.31}$ Gyr for FSR~1758. Suggesting that FSR~1758 is about $\sim$11.6 Gyr old. Clearly, deeper photometry, especially reaching below the MS turnoff, is required to best estimate the age.

The $1\sigma$  region (the shaded stripes in Figure \ref{Figure4}) is mostly affected by the uncertainties on the distance (vertical) and on the age (horizontal), given our well-determined metallicity. It is relevant to mention that in the RGB region of the CMD, an age variation could be seen as a colour displacement \citep[see Figure 2 of ][]{Souza2020}. Also, we want to highlight that our probable solutions within $1\sigma$ fit well the central part of the CMD, reinforcing that the age estimation could be a reasonable determination for FSR~1758.

It is also important to note that our detailed analysis yields an estimated reddening of E(B-V)$>0.77$, which is 0.4 mag larger than previously estimated by \citet{Barba2019}, E(B-V)$\sim0.37$. A close inspection of this field show that FSR~1758 is very near to the GCs Tonantzintla 2 \citep[Ton~2;][]{Bica1996}, and NGC~6380 \citep{Ortolani1998}, where the extinction have been found to be as large as E(B$-$V)$\sim$ 1.17 -- 1.24 \citep[see, e.g.,][]{Bica1996, Ortolani1998}. This discrepancy could be due to the fact that in \citet{Barba2019} the 2MASS photometry was employed, which is less sensitive to extinction. Thus, our Bayesian estimation of E(B-V) reveals that FSR~1758 lies in a region with higher reddening, E(B$-$V>0.76), than derived by \citet{Barba2019}, which may in fact have been erroneous on the distance estimation of FSR~1758. This value also appears to be overestimated in \citet{Barba2019}, as confirmed in Figure \ref{Figure4}, which shows that a closer distance to FSR~1758 better fits the primary sequences  (horizontal branch and RGB) of the cluster.

It is also important to mention, that our estimate distance (8.84 kpc) is in excellent agreement with the mean parallaxes reported by \citet{Vasiliev2021}, 0.119 $\pm$0.011 mas (296 stars), 8.40 kpc. It is important to note that independent methods seem to converge reasonably.

\section{Mass}
\label{section5}

The precise APOGEE-2 \textit{RV} information of our 15 sample stars were combined with other existing \textit{RV} measurements of  FSR~1758 members. In particular, \citet{Villanova2019} provided $RV$ determinations for nine additional cluster members from optical high-resolution MIKE spectra, while \citet{Simpson2019} reported \textit{RVs} for two additional cluster stars with \texttt{Gaia} radial velocities. This yields a unique collection of 26 likely members of FSR~1758 with \textit{RV} information. With this large sample in hand, we report a mean radial velocity of FSR~1758 of $+225.73\pm0.69$ km s$^{-1}$ with a velocity dispersion of  3.51$\pm$0.49 km s$^{-1}$, which is in good agreement with the mean and dispersion  reported by \citet{Villanova2019}, $+226.8\pm1.6$ km s$^{-1}$ and $4.9\pm1.2$ km s$^{-1}$.

\begin{figure}
	\begin{center}
		\includegraphics[width=90mm]{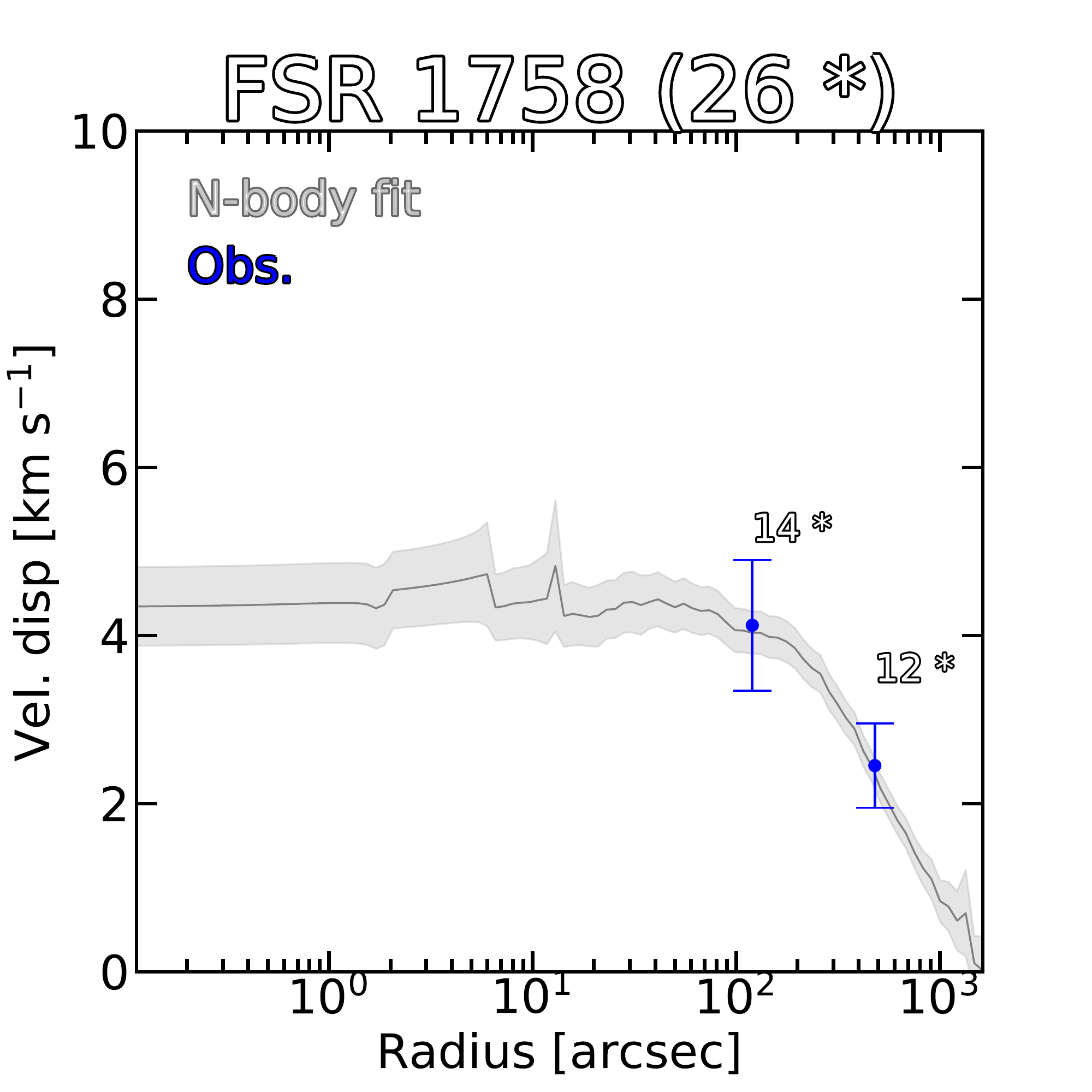}
		\caption{Line-of-sight velocity dispersion versus radius for our target cluster stars from APOGEE-2 plus \citet{Villanova2019} and \citet{Simpson2019} data set (blue dots). The error bars are refered as $\sigma_{RV}/\sqrt{2\times N}$. The prediction of the best-fitting updated (private communication) \textit{N}-body model from \citep{Baumgardt2018} and \citet{Baumgardt2019} is shown as a solid-grey line, and the light-grey shaded region indicates the 1$\sigma$ uncertainty from the fit.} 
		\label{Figure3}
	\end{center}
\end{figure}

With this relatively large sample of stars with \textit{RV} information, we match the line-of-sight dispersion profiles to the updated version of \textit{N}-body simulations (private communication with H. Baumgardt) of  FSR~1758 from \citet{Baumgardt2018, Baumgardt2019}, as shown in Figure \ref{Figure3}, and thus determine the most likely mass of the cluster from kinematic constraints. 

We adopted two radial bins (with bin centers of 2\arcmin and 8\arcmin), chosen to ensure that at least twelve stars were in each bin; resulting in the two points shown in \ref{Figure3}. We find $\sigma_{0}\sim4.3\pm0.5$ km s$^{-1}$. This yields a present-day estimated mass of $\sim2.9\pm0.6\times$10$^{5}$ M$_{\odot}$, which suggests that FSR~1758 is as massive as NGC 6752 ($\sim2.32\pm0.003 \times10^{5}$ M$_{\odot}$) \citep{Baumgardt2018, Baumgardt2019}.

\section{Orbital parameters of FSR~1758}
\label{section6}

The orbit of  FSR~1758 has been extensively discussed in \citet{Villanova2019}, \citet{Simpson2019}, and \citet{Yeh2020}, using different Galactic models. In general, the orbital solutions determined in those studies are consistent, concluding that FSR~1758 is on a highly eccentric and retrograde orbit, with apo-Galacticon $<$20 kpc  and with inner incursions close to the bulge region. 

\subsection{The Galactic model}

Here, we re-visit the orbit of  FSR~1758 by adopting the new state-of-the art orbital integration model \texttt{GravPot16}\footnote{https://gravpot.utinam.cnrs.fr} with a "boxy/peanut" bar structure in the bulge region along with other composite stellar components \citep{FT_Dynamics}, which best fit the known structural and dynamical parameters of our Galaxy, to the best of our knowledge. It is important to note that \citet{Villanova2019} employed the \texttt{GravPot16} model from the web service (\url{https://gravpot.utinam.cnrs.fr}) with a simple "prolate" bar structure.  

The structural parameters of our bar model (mass, present-day orientation and pattern speed)  are 1.1$\times$10$^{10}$ M$_{\odot}$ \citep{FT_Dynamics}, 20$^{\circ}$ \citep{Rodriguez-Fernandez2008, FT_Dynamics}, and 41 km s$^{-1}$ kpc \citep{Sanders2019}, respectively, consistent with observational estimates. The bar scale lengths are $x_0=$1.46 kpc, $y_{0}=$ 0.49 kpc, and $z_0=$0.39 kpc, and the middle region ends at the effective semimajor axis of the bar $Rc = 3.28$ kpc , whose density profile is exactly the same as in \citet{Robin2012}.

For reference, the Galactic convention adopted by this work is: $X-$axis oriented toward $l=$ 0$^{\circ}$ and $b=$ 0$^{\circ}$, $Y-$axis is oriented toward $l$ = 90$^{\circ}$ and $b=$0$^{\circ}$, and the disc rotates toward $l=$ 90$^{\circ}$; the velocity is also oriented in these directions. Following this convention, the Sun's orbital velocity vectors are [U$_{\odot}$,V$_{\odot}$,W$_{\odot}$] = [$11.1$, $12.24$, 7.25] km s$^{-1}$ \citep{Brunthaler2011}. The model has been rescaled to the Sun's galactocentric distance, 8.3 kpc, and the local rotation velocity of $244.5$ km s$^{-1}$  \citep{Sofue2015}.

\subsection{Orbital Elements}

For the computation of Galactic orbits, we have employed a simple Monte Carlo approach and the Runge-Kutta algorithm of seventh to eight order, elaborated by \citet{fehlberg68}. As input parameters we adopted the following observables: (1) \textit{RV} $=$ 225.73 km s$^{-1}$ (see Section \ref{section5}); absolute proper motions $\mu_{\alpha}\cos(\delta)=-2.85$ mas yr$^{-1}$ and $\mu_{\delta}=2.47$ mas yr$^{-1}$ (see Section \ref{fsr1758}); and a range of heliocentric distances from $8.84$ kpc (as estimated from our isochrone fits in Section \ref{section4}) and $11.5\pm1$ kpc from \citet{Barba2019}. The uncertainties in the input data (e.g. $\alpha$, $\delta$, distance, proper motions and \textit{RV} errors) were randomly propagated as 1$\sigma$ variation in a Gaussian Monte Carlo resampling. Thus, we ran ten thousand orbits, computed backwards in time during 3 Gyr. The median value of the orbital elements were found for these 10,000 realizations, with uncertainty ranges given by the 16$^{\rm th}$ and 84$^{\rm th}$ percentile. The resulting orbital elements are listed in Table \ref{Table2}. It is important to note that we list the minimal and maximum value of the z-component of the angular momentum in the inertial frame, since this quantity is not conserved in a model with non-axiymmetric structures like \texttt{GravPot16}. In the case of FSR~1758, L$_{\rm z, min}$ and L$_{\rm z, max}$ are close to each other within the uncertainties, thus confirming the genuine retrograde nature of  FSR~1758, as shown on the top-right panel in Figure \ref{Figure5}.

Overall, our orbital study based on a different Galactic model configuration than previous studies confirms that FSR~1758 lies in an radial, eccentric, and retrograde halo-like orbit, which circulates across the "bulge /bar" region with relatively high vertical excursions  from the Galactic plane, |Z|$_{\rm max}\lesssim$ 8 kpc, depending on the adopted heliocentric distance as listed in Table \ref{Table2}. 

As noted above,  there are several studies of FSR~1758 that have attempted to unveil its origin. For instance, \citet{Barba2019}, \citet{Myeong2019}, and \citet{Massari2019} proposed that it is likely part of the Sequoia dwarf galaxy, while \citet{Simpson2019} suggested that FSR~1758 is an intruder from the halo into the inner Galaxy. More recent work by \citet{Villanova2019} and \citet{Yeh2020} suggest that the chemical enrichment,  high eccentricity, and retrograde orbital configuration of FSR~1758 is not uncommon among other halo GCs, thus favoring a possible \textit{in situ} origin.

To alleviate this tension, we examine the orbital properties of FSR~1758 within the \texttt{GravPot16}$+$"boxy/peanut"-bar computational environment. Figures \ref{Figure5} and Figure \ref{Figure7} provide an "Orbital Energy" map that identifies the main structures of GCs with an accreted and in situ origin, respectively. This diagram provides one of the best dynamical representations of the conserved motion quantities in a non-axisymmetric MW model. These include the plot of the characteristic orbital energy, (E$_{\rm max}$+ E$_{\rm min}$)/2, versus the orbital Jacobi constant (E$_{\rm J}$), as envisioned by \citet{Moreno2015}, with E$_{\rm max}$, E$_{\rm min}$ the maximum and minimum energies per unit mass along each orbit, and computed with respect to the Galactic inertial frame, while E$_{\rm J}$ has a constant value in the non-inertial reference frame where the bar is at rest. 

\subsection{Comparison of the orbital elements of FSR~1758 to other globular clusters}

\begin{figure*}
	\begin{center}
		\includegraphics[width=180mm]{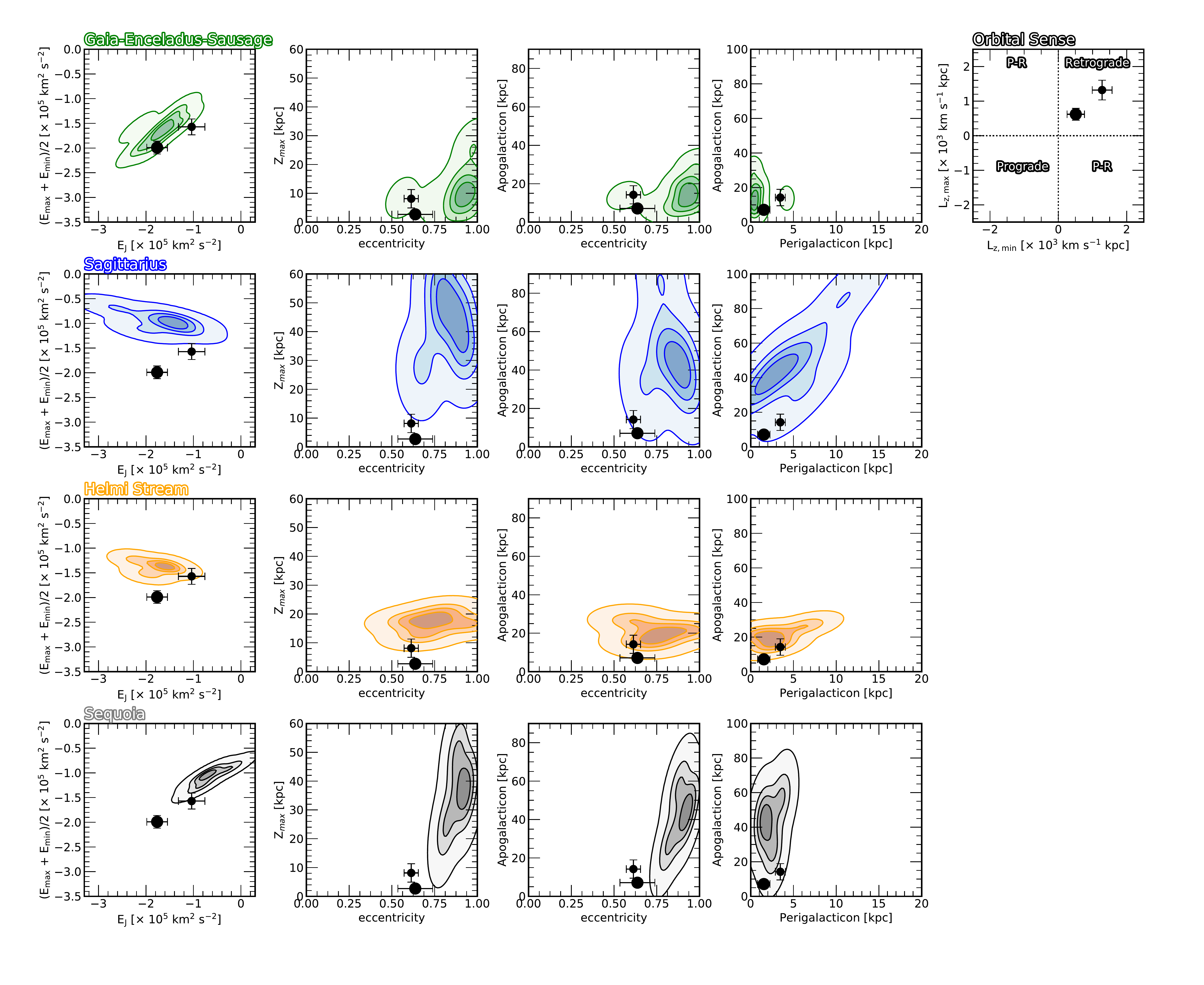}
		\caption{Kernel Density Estimation (KDE) models of the characteristic orbital energy ((E$_{\rm max}$ + E$_{\rm min}$)/2), the orbital Jacobi energy (E$_{\rm J}$), orbital pericenter and apocenter, orbital eccentricity, maximum vertical height above the Galactic plane for GCs with an accreted origin \citep[e.g.,][]{Massari2019}. FSR~1758 is highlighted with black dot symbols, by considering a heliocentric distance of 8.84 kpc (large symbol), and 11.5 kpc (small symbol) The top-right panel show the minimal and maximum value of the z-component of the angular momentum in the inertial frame, and indicates the regions dominated by prograde and retrograde orbits, and those dominated by orbits that change their sense of motion from prograde to retrograde (P-R).} 
		\label{Figure5}
	\end{center}
\end{figure*}

\begin{figure*}
	\begin{center}
		\includegraphics[width=180mm]{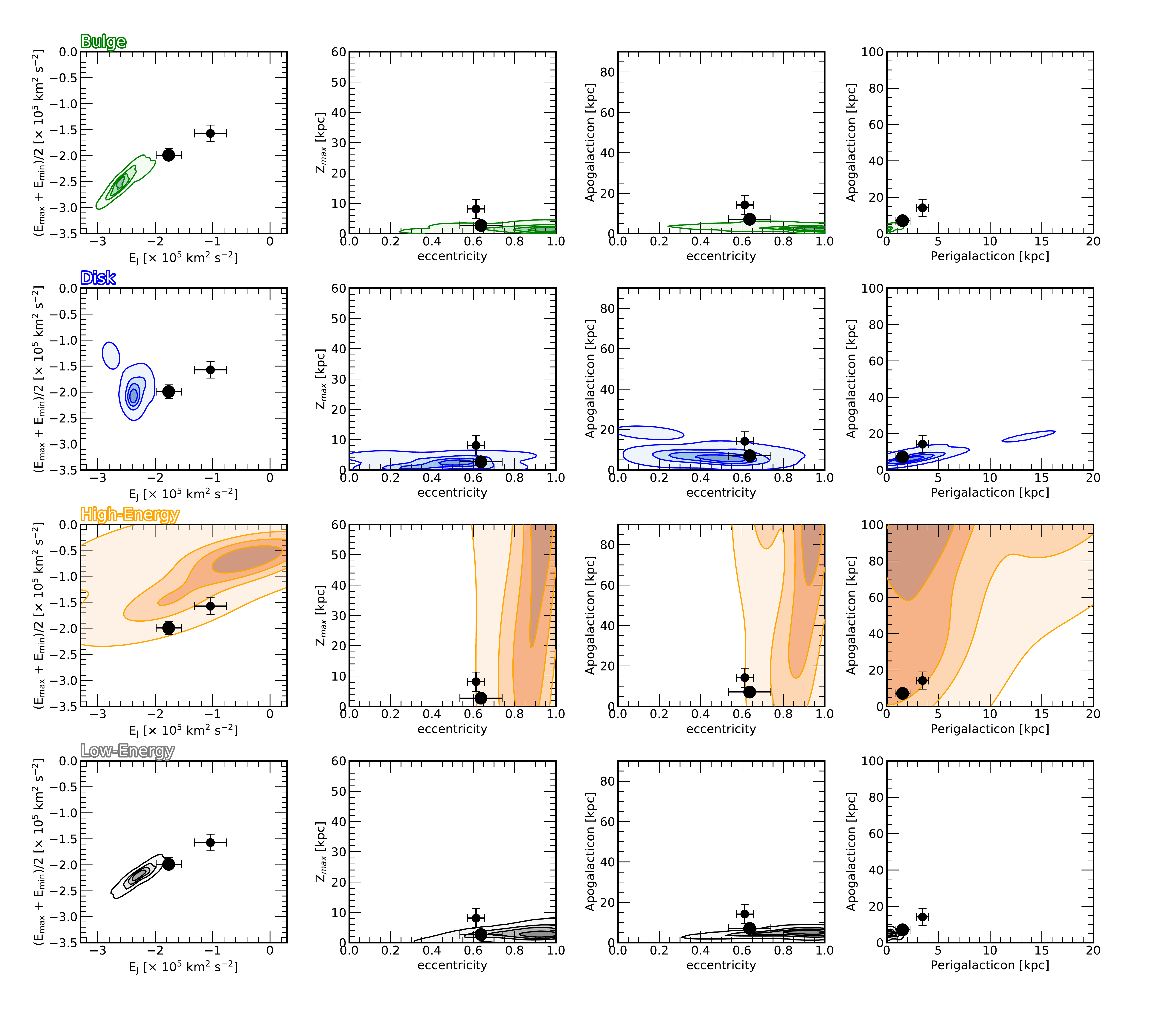}
		\caption{Same as Figure \ref{Figure5}, but now considering GCs with an in situ origin \citep{Massari2019}.} 
		\label{Figure7}
	\end{center}
\end{figure*}

We computed the above quantities for the recent compilation of GCs listed in \citet{Vasiliev2021},  adopting the same Monte Carlo approach as described above. Further, we adopted the same GC classification as presented in \citet{Massari2019} to identify the main structures in our maps, and computed the Kernel Density Estimation (KDE) models for each of  the GCs groups belonging to the Main-Disc, Main-Bulge, High-Energy, Low-Energy, Sequoia, Gaia-Enceladus-Sausage, Helmi Stream, and Saggitarius population. The resulting KDE models and the their corresponding structure are presented in Figures \ref{Figure5} and \ref{Figure7}. This complementary analysis provides us with important insight into the origin of FSR~1758. 

 Figures \ref{Figure5} and \ref{Figure7} reveal that FSR~1758 lies in the area of this diagram mostly dominated by  populations with retrograde orbits. In this diagram, FSR~1758 is located on the boundary of four main structures, the Gaia-Enceladus-Sausage, Helmi Stream, and Sequoia accretion event, and the structure dominated by High-Energy GCs. However, a close inspection of other orbital elements (the orbital eccentricity, the vertical excursions above the Galactic plane, $Z_{max}$, and the apo-/peri-galactic distance), as presented in Figures~\ref{Figure5} and \ref{Figure7},  reveal some significant differences. Figure~\ref{Figure7} shows that the orbital elements of the FSR~1758 barely overlap the tail of the distribution of the High-Energy structure, which is mostly dominated by highly eccentric and radial orbits with extremely large vertical excursions above the Galactic plane, making less probable that FSR~1758 belongs to this structure. 
 
 Figure \ref{Figure5} shows that the Sagittarius structure is dominated by GCs in highly eccentric and radial orbits with $Z_{max}\gtrsim10 $ kpc, barely overlapped by the orbit of FSR~1758 and far away in the ``Orbital Energy" map. The same figure reveals that, depending on the adopted heliocentric distance, FSR~1758 departs relatively far away in the ``Orbital Energy" map from the Sequoia structure, and is beyond the tail in the eccentricity--$Z_{plane}$, and barely overlapping the tail of other orbital elements. In other words, GCs in the Sequoia structure are dominated by mostly highly eccentric orbits with large vertical excursions from the Galactic plane ($\gtrsim 20$ kpc), and also reaching large apogalactocentric distances ($>30$ kpc), making FSR~1758 a low-probability candidate for association with the Sequoia accretion event. This is also supported by the very old age estimate provided in Section \ref{section4}. 
 
 In addition, Figure \ref{Figure5} reveals that the Gaia-Enceladus-Sausage structure is more confined to the Galactic plane ($< 20$ kpc), with apogalactocentric distances within $<30$ kpc and eccentricities as low as $e \gtrsim 0.55$). Thus, FSR~1758 exhibits orbital properties overlapping fairly well to all the orbital elements of the Gaia-Enceladus-Sausage structure. Therefore, our analysis favors a possible link of FSR~1758 with the Gaia-Enceladus-Sausage accretion event rather than to the Sequoia structure or any possible \textit{in situ} origin. We conclude that it is possible that FSR~1758 has likely been dynamically ejected into the inner halo of the MW from the Gaia-Enceladus-Sausage dwarf galaxy.

\begin{table}
	\begin{center}
		\setlength{\tabcolsep}{1.5mm}  
		\caption{Main orbital properties of FSR~1758}
		\begin{tabular}{ll}
			 \hline
             \hline
              $d_{\odot} = $ 8.84 kpc (This work) & \\
             \hline
             \hline
			Perigalacticon          &  1.53$\pm$0.72 kpc \\			
			Apogalacticon           &  7.11$\pm$ 1.24 kpc \\
			eccentricity               &  0.64$\pm$0.10 kpc \\
			|Z|$_{\rm max}$   & 2.71$\pm$0.82 kpc \\
			
			L$_{\rm z, min}$   &  0.51$\pm$0.26 $\times{}10^3 $  km s$^{-1}$  kpc \\
			L$_{\rm z, max}$  &  0.62$\pm$0.17 $\times{}10^3 $  km s$^{-1}$  kpc \\
			E$_{\rm J}$             &  $-$1.76$\pm$0.22 $\times{}10^5 $  km$^{2}$ s$^{2}$\\
			E$_{\rm min}$        & $-$2.09$\pm$0.16 $\times{}10^5 $  km$^{2}$ s$^{2}$ \\
			E$_{\rm max}$       &  $-$1.88$\pm$0.10$\times{}10^5 $  km$^{2}$ s$^{2}$ \\
			 \hline
			 \hline
$d_{\odot} = $11.5 kpc \citep{Barba2019}& \\
\hline
\hline
Perigalacticon          &  3.45$\pm$0.6 kpc \\
Apogalacticon           &  14.2$\pm$ 4.7 kpc \\
eccentricity               &  0.61$\pm$0.04 kpc \\
|Z|$_{\rm max}$   & 8.1$\pm$3.2 kpc \\
L$_{\rm z, min}$   &  0.12$\pm$0.03 $\times{}10^3 $  km s$^{-1}$  kpc \\
L$_{\rm z, max}$  &  0.13$\pm$0.03 $\times{}10^3 $  km s$^{-1}$  kpc \\
E$_{\rm J}$             &  $-$1.04$\pm$0.28 $\times{}10^5 $  km$^{2}$ s$^{2}$\\
E$_{\rm min}$        & $-$1.59$\pm$0.18 $\times{}10^5 $  km$^{2}$ s$^{2}$ \\
E$_{\rm max}$       &  $-$1.56$\pm$0.14 $\times{}10^5 $  km$^{2}$ s$^{2}$ \\
\hline 			 
		\end{tabular}  \label{Table2}
	\end{center}
\end{table}   

\section{Summary and concluding remarks}
\label{section7}

We have performed the first high-resolution (R$\sim$22,000) spectral analysis in the \textit{H}-band of the intriguing GC FSR~1758, which was recently discovered toward the bulge region. We used APOGEE data from the \texttt{CAPOS} survey of fifteen RGB stars which are all very likely members.  We also provided a dynamical study of the cluster by adopting the state-of-the-art \textit{GravPot16} Galactic model with a "boxy/peanut" bar structure, and employing newly revised estimation of its parameters, inlcuding PMs, \textit{RV} and distance. 

We found a mean metallicity of [Fe/H]$=-1.36$ and [Fe/H]$=-1.43$, by adopting photometric and spectroscopic atmospheric parameters for FSR~1758, respectively. This value is somewhat more metal rich than previously reported \citep[see, e.g.,][]{Barba2019, Villanova2019}, with no evidence for an intrinsic metallicity spread. The mean radial velocity of the cluster, determined from a large sample (26 stars), was found to be $+225.73\pm0.69$ km s$^{-1}$ with a velocity dispersion of 3.51 km s$^{-1}$, and the mean PMs in \texttt{Gaia} EDR3 are found to be $\mu_{\alpha}\cos(\delta)=-2.85\pm0.05$ mas yr$^{-1}$ and $\mu_{\delta}=2.47\pm0.05$ mas yr$^{-1}$. These values are in good agreement with the literature within the uncertainties. 

We derived detailed abundances for 10 species, several of which were not studied previously. Thus, C, N, and Si abundances are reported for the first time. We found evidence for a C-N anti-correlation and N-Al correlation of the sample analyzed. These signatures of multiple populations support the findings from the  Na-O anti-correlation found by \citet{Villanova2019} that this GC displays this phenomenon. All of the evidence,  including multiple populations, no metallicity spread, relatively high metallicity and small velocity dispersion, clearly indicate that FSR~1758 is a bonafide GC, and not the remnant nucleus of a dwarf galaxy, as originally speculated by \citet{Barba2019}. Note that the suspected tidal tails they noted are now known to be simply outliers in the field star PM distribution. However, this object remains as an outlier in the size vs. Galactocentric distance plot, being very large for a GC so near the center, undoubtedly stemming from the fact that its orbit only brings it this close on occasion.

The \texttt{SIRIUS} code is applied to provide a self-consistent  age of FSR~1758 by using a statistical isochrone fitting to the \textit{Gaia} band system. We find that FSR~1758 is as old as 11.6$^{+1.25}_{-1.31}$ Gyr. 

Additionally, the new dataset was combined with other existing \textit{RV} measurements, allowing a better constraint on the predictions of the \textit{N}-body line-of-sight velocity dispersion of FSR~1758. This yields an estimated mass of $\sim$2.9$\pm$0.6 $\times$ 10$^{5}$ M$_{\odot}$, which make FSR~1758 as massive as NGC 6752. We also confirm the prevalence of a small velocity dispersion ($3.51\pm0.49$ km s$^{-1}$), which is typical for a GC. 

The orbital properties of FSR~1758, combined with its chemical properties and old age estimation, leads to the conclusion that FSR~1758 matches well with the physical properties seen in a good fraction of GCs associated with the Gaia-Enceladus-Sausage accretion event, rather than those associated with the Sequoia event. We  favor the association of FSR~1758 with the Gaia-Enceladus-Sausage accretion event, which is also consistent with the age-metallicity relation provided by \citet{Massari2019}. Thus, the "Sequoia" GC is not in fact a member of the dwarf galaxy progenitor  whose name it inspired \citep{Myeong2019, Koppelman2019}.
	
	\begin{acknowledgements}  
		The author is grateful for the enlightening feedback from the anonymous referee. We warmly thank Holger Baumgardt for providing his more recent numerical \textit{N}-body modeling of the line-of-sight velocity dispersion of FSR~1758.  M. I. R.-C. was supported by VRIIP, UA through  "Concurso de Asistentes de Investigaci\'on" funded by  "Ministerio de Educaci\'on Chile" No. ANT1855 and ANT1856 projects and gratefully acknowledges support from the Graduate School of the Universidad de Antofagasta for their support with "Beca de Excelencia" (Scholarship of Excellence). D.G. gratefully acknowledges support from the Chilean Centro de Excelencia en Astrof\'isica y Tecnolog\'ias Afines (CATA) BASAL grant AFB-170002. D.G. also acknowledges financial support from the Direcci\'on de Investigaci\'on y Desarrollo de la Universidad de La Serena through the Programa de Incentivo a la Investigaci\'on de Acad\'emicos (PIA-DIDULS). D.G. and D.M. gratefully acknowledge support from the Chilean Centro de Excelencia en Astrof\'isica y Tecnolog\'ias Afines (CATA) BASAL grant AFB-170002. D.M. is also supported by Fondecyt 1170121. T.C.B. acknowledges partial support from grant PHY 14-30152, Physics Frontier Center/JINA Center for the Evolution of the Elements (JINA-CEE), awarded by the US National Science Foundation. SOS acknowledges the FAPESP PhD fellowship 2018/22044-3. SOS and APV acknowledge the DGAPA-PAPIIT grant IG100319. BB acknowledges grants from FAPESP, CNPq and CAPES - Financial code 001. J.A.-G. acknowledges support from Fondecyt Regular 1201490 and from ANID -- Millennium Science Initiative Program -- ICN12\_009 awarded to the Millennium Institute of Astrophysics MAS. L.H. gratefully acknowledges support  provided by National Agency for Research and Development (ANID)/CONICYT-PFCHA/DOCTORADO NACIONAL/2017-21171231. \\

        This work has made use of data from the European Space Agency (ESA) mission \textit{Gaia} (\url{http://www.cosmos.esa.int/gaia}), processed by the \textit{Gaia} Data Processing and Analysis Consortium (DPAC, \url{http://www.cosmos.esa.int/web/gaia/dpac/consortium}). Funding for the DPAC has been provided by national institutions, in particular the institutions participating in the \textit{Gaia} Multilateral Agreement.\\

        Funding for the Sloan Digital Sky Survey IV has been provided by the Alfred P. Sloan Foundation, the U.S. Department of Energy Office of Science, and the Participating Institutions. SDSS- IV acknowledges support and resources from the Center for High-Performance Computing at the University of Utah. The SDSS web site is www.sdss.org. SDSS-IV is managed by the Astrophysical Research Consortium for the Participating Institutions of the SDSS Collaboration including the Brazilian Participation Group, the Carnegie Institution for Science, Carnegie Mellon University, the Chilean Participation Group, the French Participation Group, Harvard-Smithsonian Center for Astrophysics, Instituto de Astrof\`{i}sica de Canarias, The Johns Hopkins University, Kavli Institute for the Physics and Mathematics of the Universe (IPMU) / University of Tokyo, Lawrence Berkeley National Laboratory, Leibniz Institut f\"{u}r Astrophysik Potsdam (AIP), Max-Planck-Institut f\"{u}r Astronomie (MPIA Heidelberg), Max-Planck-Institut f\"{u}r Astrophysik (MPA Garching), Max-Planck-Institut f\"{u}r Extraterrestrische Physik (MPE), National Astronomical Observatory of China, New Mexico State University, New York University, University of Notre Dame, Observat\'{o}rio Nacional / MCTI, The Ohio State University, Pennsylvania State University, Shanghai Astronomical Observatory, United Kingdom Participation Group, Universidad Nacional Aut\'{o}noma de M\'{e}xico, University of Arizona, University of Colorado Boulder, University of Oxford, University of Portsmouth, University of Utah, University of Virginia, University of Washington, University of Wisconsin, Vanderbilt University, and Yale University.\\
	\end{acknowledgements}



\end{document}